\documentclass[a4paper,11pt]{article}
\pdfoutput=1 

\usepackage{jheppub} 

\usepackage{natbib}
\usepackage{epsfig}
\usepackage{graphicx}
\usepackage{dcolumn}
\usepackage{amsmath}
\usepackage{amssymb}
\usepackage{enumerate}
\usepackage{subfigure}    
\usepackage{multirow}
\usepackage{color} 

\newcommand{\ssthac}{\sin^{2}2\theta_{13}}

\newcommand{\dcp}{\delta_{CP}}
\newcommand{\nue}{\nu_{e}}
\newcommand{\numu}{\nu_{\mu}}

\newcommand{\anue}{\bar{\nu}_{e}}
\newcommand{\anumu}{\bar{\nu}_{\mu}}

\newcommand{\pyha}{Pyh\"asalmi }

\newcommand{\refeq}[1]{Eq.~\ref{#1}}
\newcommand{\refsec}[1]{Section~\ref{#1}}

\newcommand{\reftab}[1]{Tab.~\ref{#1}}

\newcommand{\reffig}[1]{Fig.~\ref{#1}}

\title{The LBNO long-baseline oscillation sensitivities with two conventional neutrino beams at different baselines}

\author[o]{S.K.\,Agarwalla,}
\author[a]{L.\,Agostino,}
\author[u]{M.\,Aittola,}
\author[b]{A.\,Alekou,}
\author[x]{B.\,Andrieu,}
\author[b]{F.\,Antoniou,}
\author[aa]{R.\,Asfandiyarov,}
\author[y]{D.\,Autiero,}
\author[k]{O.\,B\'esida,}
\author[r]{A.\,Balik,}
% ...

\affiliation[a]{APC, AstroParticule et Cosmologie, Universit\'e Paris Diderot, CNRS/IN2P3, CEA/Irfu, Observatoire de Paris, Sorbonne Paris Cit\'e, 10, rue Alice Domon et L\'eonie Duquet, 75205 Paris Cedex 13, France}
\affiliation[b]{CERN, Geneva, Switzerland}
% ...
\affiliation[k]{IRFU, CEA Saclay, Gif-sur-Yvette, France}
% ...
\affiliation[o]{Institute of Physics, Sachivalaya Marg, Sainik School Post, Bhubaneswar 751005, India}
% ...
\affiliation[r]{LAPP, Universit\'e de Savoie, CNRS/IN2P3, F-74941 Annecy-le-Vieux, France}
% ...
\affiliation[u]{Oulu Southern Institute and Department of Physics, University of Oulu, Finland}
% ...
\affiliation[x]{UPMC, Universit\'e Paris Diderot, CNRS/IN2P3, Laboratoire de Physique Nucl\'eaire et de Hautes Energies (LPNHE), Paris, France}
\affiliation[y]{Universit\'e de Lyon, Universit\'e Claude Bernard Lyon 1, IPN Lyon (IN2P3), Villeurbanne, France}
% ...
\affiliation[aa]{University of Geneva, Section de Physique, DPNC, Geneva, Switzerland}
% ...

\abstract{
The proposed Long Baseline Neutrino Observatory (LBNO) initially consists of 
$\sim 20$~kton liquid double phase TPC complemented by a magnetised
iron calorimeter, to be installed at the \pyha mine, at
a distance of 2300 km from CERN. The conventional neutrino beam is produced
by 400~GeV protons accelerated at the SPS accelerator delivering 700~kW
of power. 
The long baseline provides a unique 
opportunity to study neutrino flavour oscillations over their 
1st and 2nd oscillation maxima exploring the $L/E$ behaviour,
and distinguishing effects arising from $\delta_{CP}$ and matter.
In this paper we show how this comprehensive physics case can be further
enhanced and complemented if a neutrino beam produced at the Protvino
IHEP accelerator complex, at a distance of 1160~km, and with modest power of 450~kW  is aimed towards the same far detectors. 
We show that the coupling of two independent sub-MW conventional neutrino and antineutrino beams at different baselines
from CERN and Protvino will allow to measure
CP violation in the leptonic sector at a confidence level of at least $3\sigma$ for 50\% of the true values of
$\dcp$ with a 20 kton detector.
With a far detector of 70~kton, the combination
allows a $3\sigma$ sensitivity for 75\% of the true values of $\dcp$ after
10 years of running. Running two independent neutrino beams, each at a power below 1~MW,  is more
within today's state of the art than the long-term operation of a new single high-energy multi-MW facility, which has
several technical challenges and will likely require a learning curve.}

\begin{document}
\maketitle

\section{Introduction}
Neutrino masses and oscillations are, to this day, the only
experimentally established evidence of physics 
Beyond the Standard Model (BSM).
In the three neutrino framework the oscillations of massive neutrinos
are described via
the Pontecorvo-Maki-Nakagawa-Sakata
(PMNS~\cite{Maki:1962mu,Pontecorvo:1967fh}) matrix.
This $3\times 3$ unitary matrix $U$ is generally
parameterized in terms of the three mixing angles $\theta_{12}$, $\theta_{13}$, $\theta_{23}$, and
the CP violating Dirac phase $\delta_{CP}$ (neglecting Majorana phases).
The parameter $\delta_{CP}$ is the  phase that controls the $CP$
asymmetry. 

In addition to the known $\theta_{12}$ and $\theta_{23}$, the recent measurement of the 
last mixing angle $\theta_{13}$~\cite{An:2012eh,Abe:2013hdq}, opened
the way to a new generation of long-baseline neutrino oscillation
experiments. The major goals of future long-baseline experiments such as the proposed 
LBNO~\cite{Stahl:2012exa,Agarwalla::2013kaa}, LBNE~\cite{Adams:2013qkq,Bass:2013vcg} and HyperKamiokande~\cite{Kearns:2013lea} 
are the conclusive determination of the mass hierarchy (MH) and the search for leptonic CP-violation (CPV) via
the determination of $\delta_{CP}$. 

The strategy of LBNO is the outcome of the extensive feasibility 
studies LAGUNA and LAGUNA-LBNO which lasted 6 years~\cite{Patzak:2012rz,Rubbia:2013zqa}.
LBNO aims at the construction of a double phase liquid argon neutrino observatory~\cite{Rubbia:2009md} 
complemented by a magnetised muon
detector (MIND)~\cite{Abe:2007bi}, located deep-underground at the \pyha mine.
The long-baseline neutrino oscillation programme uses a wide-band
beam from the CERN SPS accelerator at a distance of
2300~km.

In  LBNO with a baseline of 2300~km,  the $L/E$ dependence of the $\numu\rightarrow\nue$ and
of the $\anumu\rightarrow\anue$ oscillation probabilities and the very
long neutrino path through Earth allow to address both fundamental questions,
the  ordering of neutrino mass eigenstates and CP violation in the lepton sector,
in clean experimental conditions. This can be achieved by profiting from the ability
to reverse the focusing horns polarity and from the well controlled
fluxes typical of neutrino produced by accelerators. 
Its physics potential 
has been extensively evaluated previously~\cite{Stahl:2012exa,Agarwalla::2013kaa,Agarwalla::2014tca}.

It was shown that
a few years of running with the CERN SPS at 700~kW beam power and a 20~kton far detector mass
provides a direct and guaranteed discovery of MH ($>~5\sigma$) over the full phase space
of oscillation parameters, and a significance for CPV above $3\sigma$
for $\sim~25(40)\%$ of the $\dcp$ values under the expectation that
$\ssthac$ will be known with a precision of $\pm10(2.5)\%$. The setup with a 2300~km baseline allows
to optimise the conventional neutrino flux to a broad band beam covering both 1st and 2nd
maximum optimising the sensitivity for CPV and is less sensitive to the assumed systematic errors,
compared to the LBNE and HyperKamiokande setups~\cite{Agarwalla::2013kaa}.

\begin{figure}[h]
\begin{center}
\includegraphics[height = 0.49\textwidth]{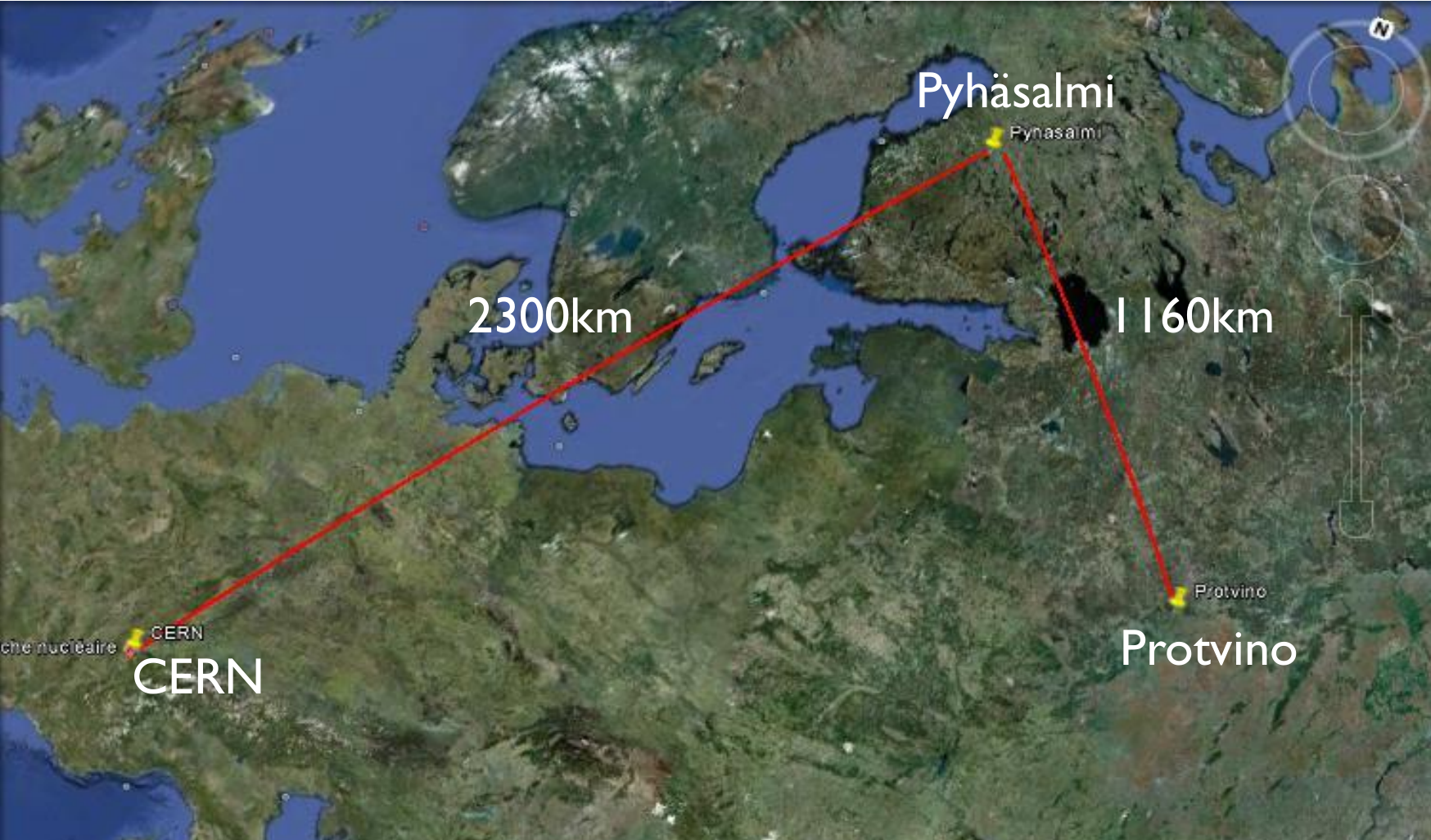}\\
\caption{View of the dual beamline configuration: CERN to \pyha (2300
  km) and Protvino to \pyha (1160 km).}
\label{fig:viewbeams}
\end{center}
\end{figure}

In this paper, we  present the physics potential of LBNO if the setup were 
further coupled to an additional neutrino beam from the Protvino
IHEP facility at a distance of 1160~km (See \reffig{fig:viewbeams}). There, an existing  proton synchrotron 
accelerator up to 70 GeV could be upgraded to reach a modest beam power of
400-450 kW and produce a neutrino beam aimed to the \pyha
mine. 
We show that the physics case for a long baseline neutrino
beam can be strengthened by a dual beam setup. 
There is a great experience worldwide, especially at the CERN SPS,  in running conventional neutrino beams with powers
ranging from 200-400~kW. Within the next decade, the NuMI beam at FNAL is planned to operate
up to 700~kW~\cite{Paley:2013sta}.
Running two independent neutrino beams, each at a power below 1~MW,  is more practical and
within known domains of experience than the long-term operation of a new single high-energy multi-MW facility, which has
several technical challenges and will likely require a learning curve. Therefore the approach discussed
in this paper with two sub-MW conventional neutrino beams would have several technical advantages
and rely on existing and proven neutrino beam technologies.  

The structure of this paper is as follows. In \refsec{sec:physics} we
review how the baseline choice influences the physics reach of a 
long-baseline experiment and which advantages a dual beam facility offers. In
\refsec{sec:protvino}, we introduce the Protvino facility and
discuss the beam specifications at this site. We present the details of our
simulations and analyses in \refsec{sec:anameth}, and our results will be shown in
\refsec{sec:results}.


\section{Choice of the baseline distance\label{sec:physics}}
A main parameter in a long-baseline experiment is
the choice of the baseline. The distance between the source of the neutrinos and the detector 
influences the physics potential of the facility in a
number of ways. In the setup considered in this paper we studied two independent and
different baselines: (1) the CERN to \pyha beam
(C2P) with 2300~km and (2) the Protvino to
\pyha beam (P2P) with 1160~km.

\subsection{CP violation sensitivity}
A well known property of the neutrino oscillation phenomenon is that
the CP asymmetry increases from the first to higher orders oscillation
maxima (see e.g.~\cite{Coloma:2011pg}).
This is understood by the fact that asymmetry due to CP has an envelope determined by~\cite{Arafune:1997hd}:
\begin{eqnarray}
\frac{2 s_{\delta}  c_{12}
 s_{12}}{   s_{13} }\cot\theta_{23}
 \frac{\Delta m^2_{21} L}{2 E}
\end{eqnarray}
which grows as a function of $L/E$. The 2nd maximum (the one at lower energy) is hence more sensitive to CPV
than the 1st maximum and an experiment able to access the
information from the 2nd maximum will have a better sensitivity to $\delta_{CP}$.

\begin{figure}[ht]
\begin{center}
\begin{tabular}{cc}
\includegraphics[clip,trim=40 40 40 40,height = 0.36\textwidth]{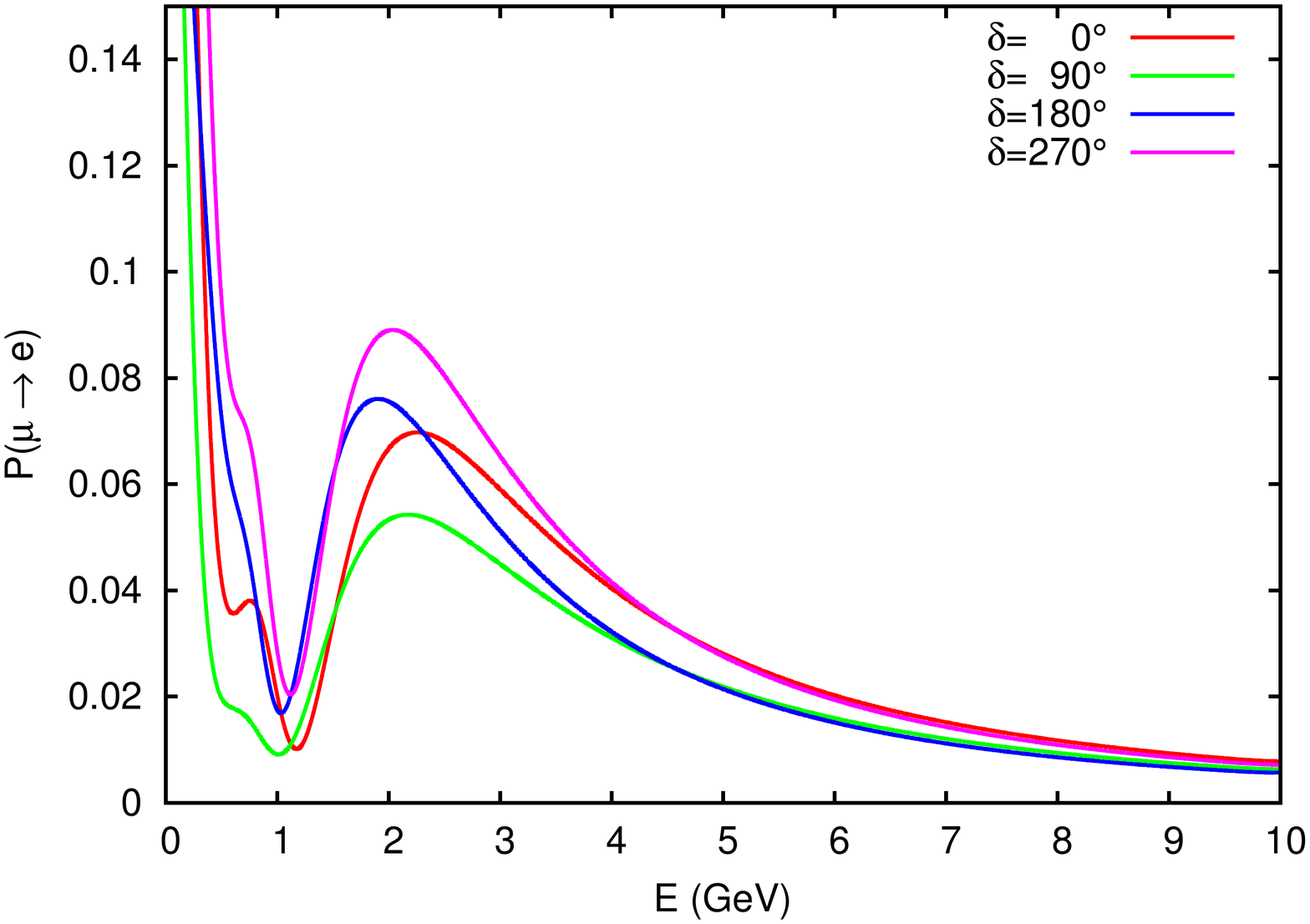}
& 
\includegraphics[clip,trim=40 40 40 40,height = 0.36\textwidth]{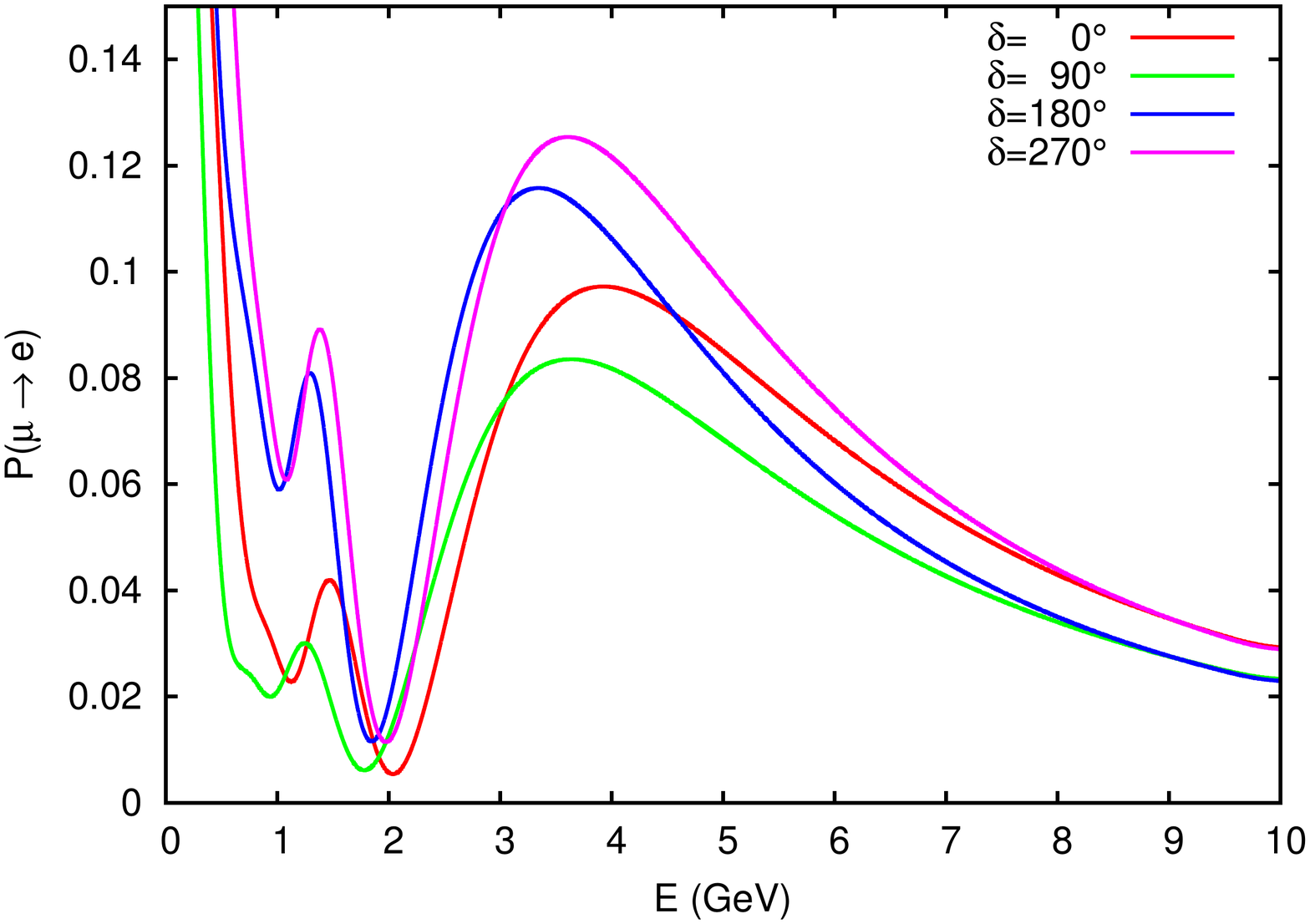}\\[-3mm]
\includegraphics[clip,trim=40 40 40 40,height = 0.36\textwidth]{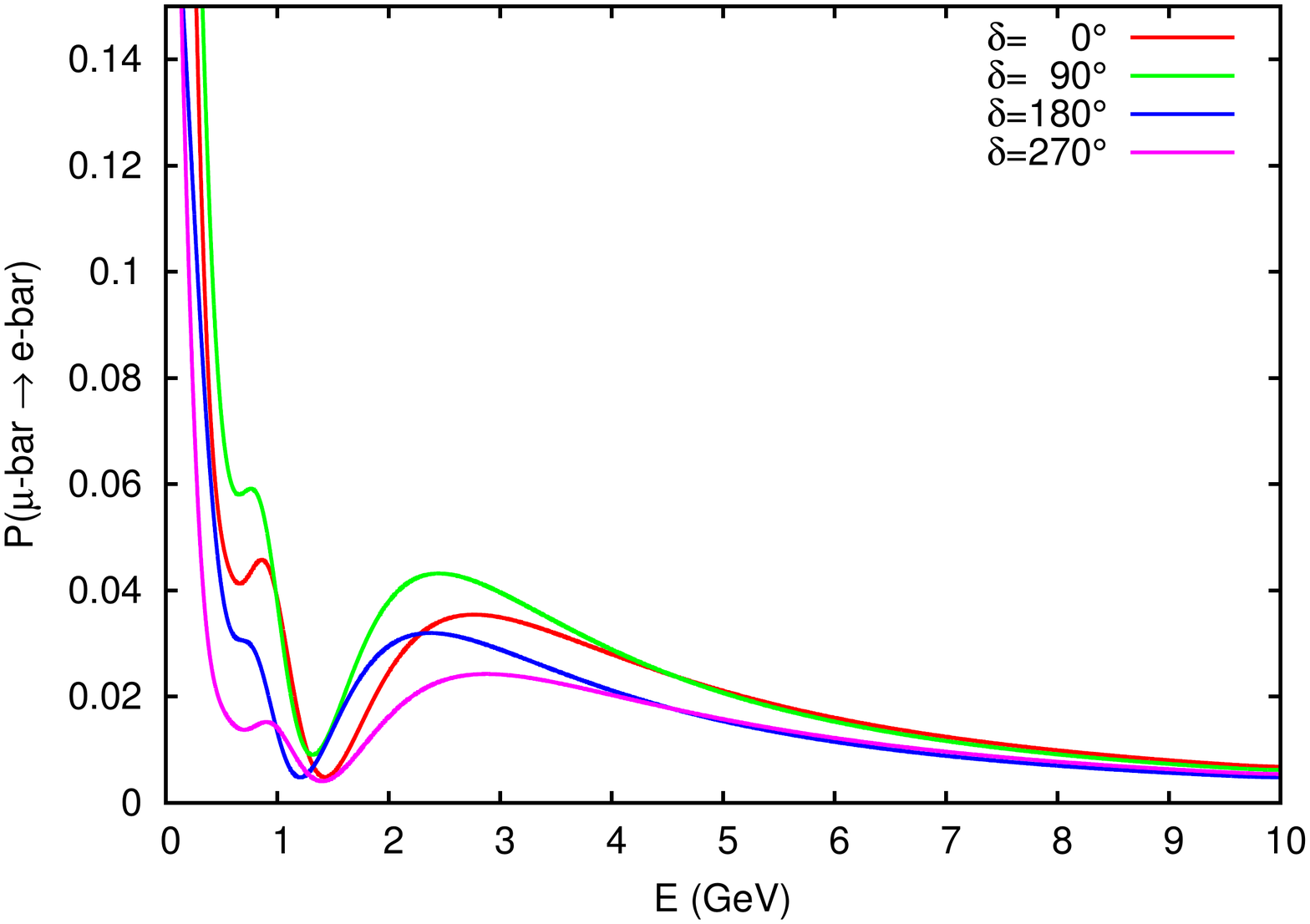}
& 
\includegraphics[clip,trim=40 40 40 40,height = 0.36\textwidth]{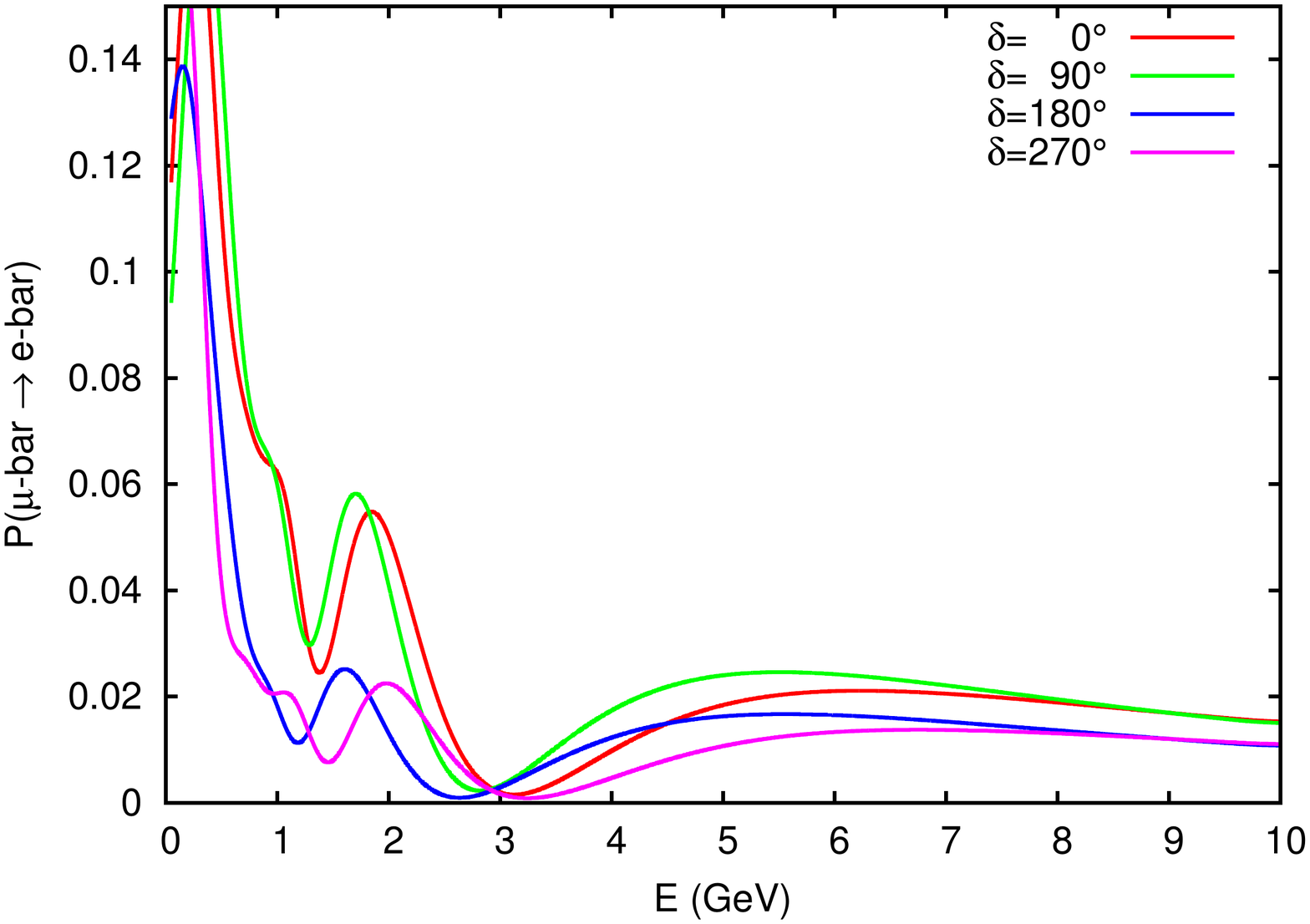}\\
a) Protvino beam & b) CERN beam
\end{tabular}
\caption{Top row (bottom row): the $\numu$ to $\nue$ ($\anumu$ to $\anue$) oscillation probability at the Pyh\"{a}salmi 
site for a selection of true values of $\dcp$. These plots assume normal 
mass ordering and have a Gaussian smear applied, as discussed in the text.}
\label{fig:prob_delta}
\end{center}
\end{figure}

The oscillation probability for P2P and C2P 
are shown in \reffig{fig:prob_delta}. 
In the plots, the probabilities have a Gaussian smearing applied to
the neutrino energy in order to
illustrate the effects of detector resolution. 
Looking at the probabilities, one can infer that spectral regions of interest are 
around the 1st and the 2nd maximum, and the regions in-between.
For P2P the second maximum region appears at rather low energies of about 0.5-1~GeV. 
For C2P due to its longer baseline, the second maximum occurs between 
1-2 GeV. 
The technical limitations of conventional neutrino beams are well understood and 
with current technological limits on magnetic field strengths and volumes, they
are not efficient at focusing hadrons with energies below $\sim 1$~GeV~\cite{Bass:2013vcg}. 
Access to the 2nd oscillation maximum is therefore more easily achieved with longer baselines.
Taking into account vanishing neutrino cross-sections at low energies (in particular for antineutrinos),
the measurement of the 2nd oscillation maximum requires in practice a baseline greater than 
$1500$~km~\cite{Agarwalla::2013kaa}. The $\nu_\mu\rightarrow \nu_\tau$~CC events where $\tau\rightarrow e\nu\nu$, 
which become more important at longer baselines due to the higher energy tails of the neutrino flux, 
act a priori as a background to electron appearance. But, as shown in Refs.~\cite{Stahl:2012exa,Agarwalla::2013kaa},
they can be kinematically separated exploiting the
excellent kinematic reconstruction of liquid argon detectors. 
In conclusion, the longer baseline C2P option is more
favorable for the observation of events in the second maximum, but both
P2P and C2P populate efficiently the region of the first maximum.

\subsection{Mass hierarchy determination sensitivity}

\begin{figure}[ht]
\begin{center}
\begin{tabular}{cc}
\includegraphics[height = 0.375\textwidth]{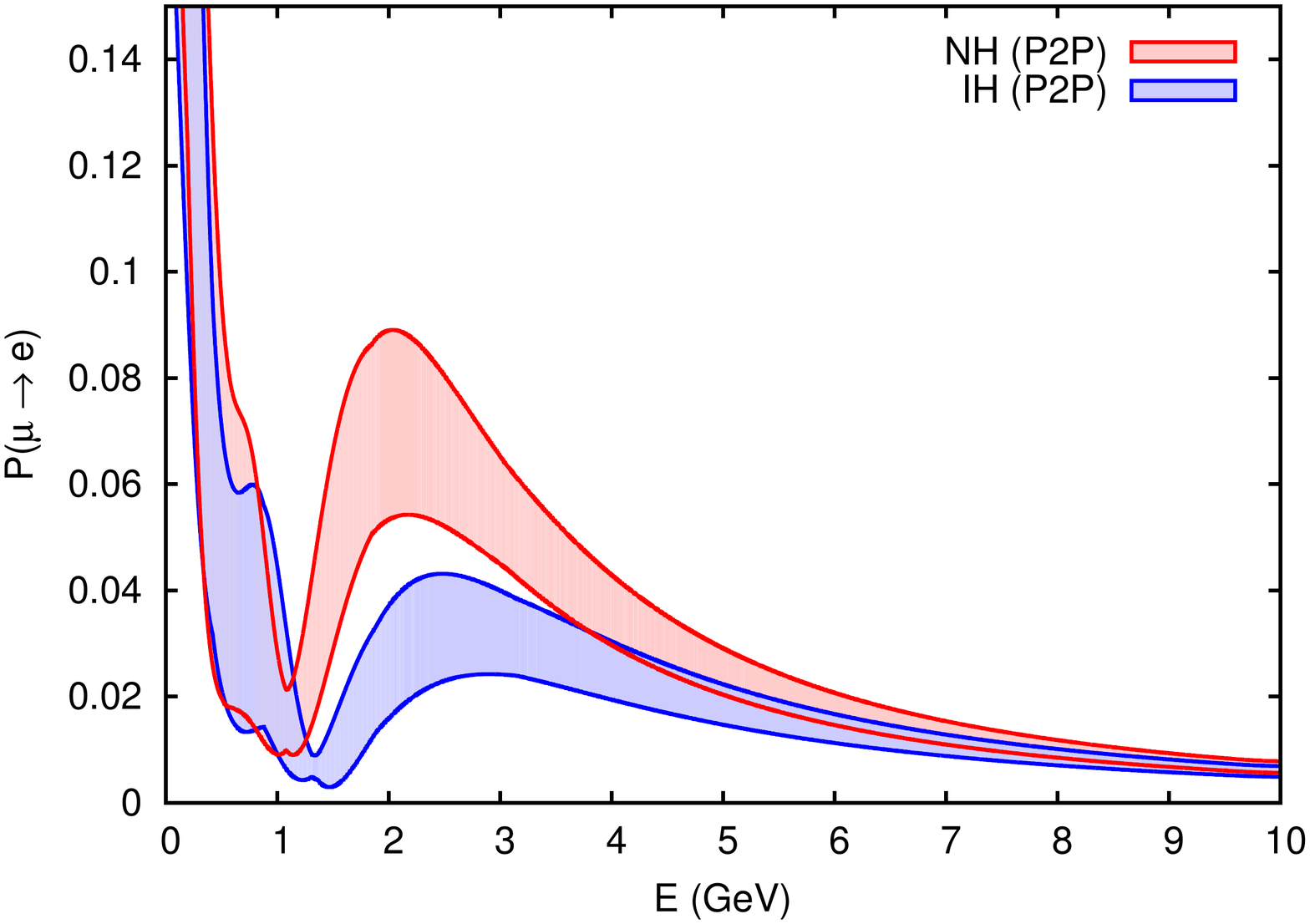}
&
\includegraphics[height = 0.375\textwidth]{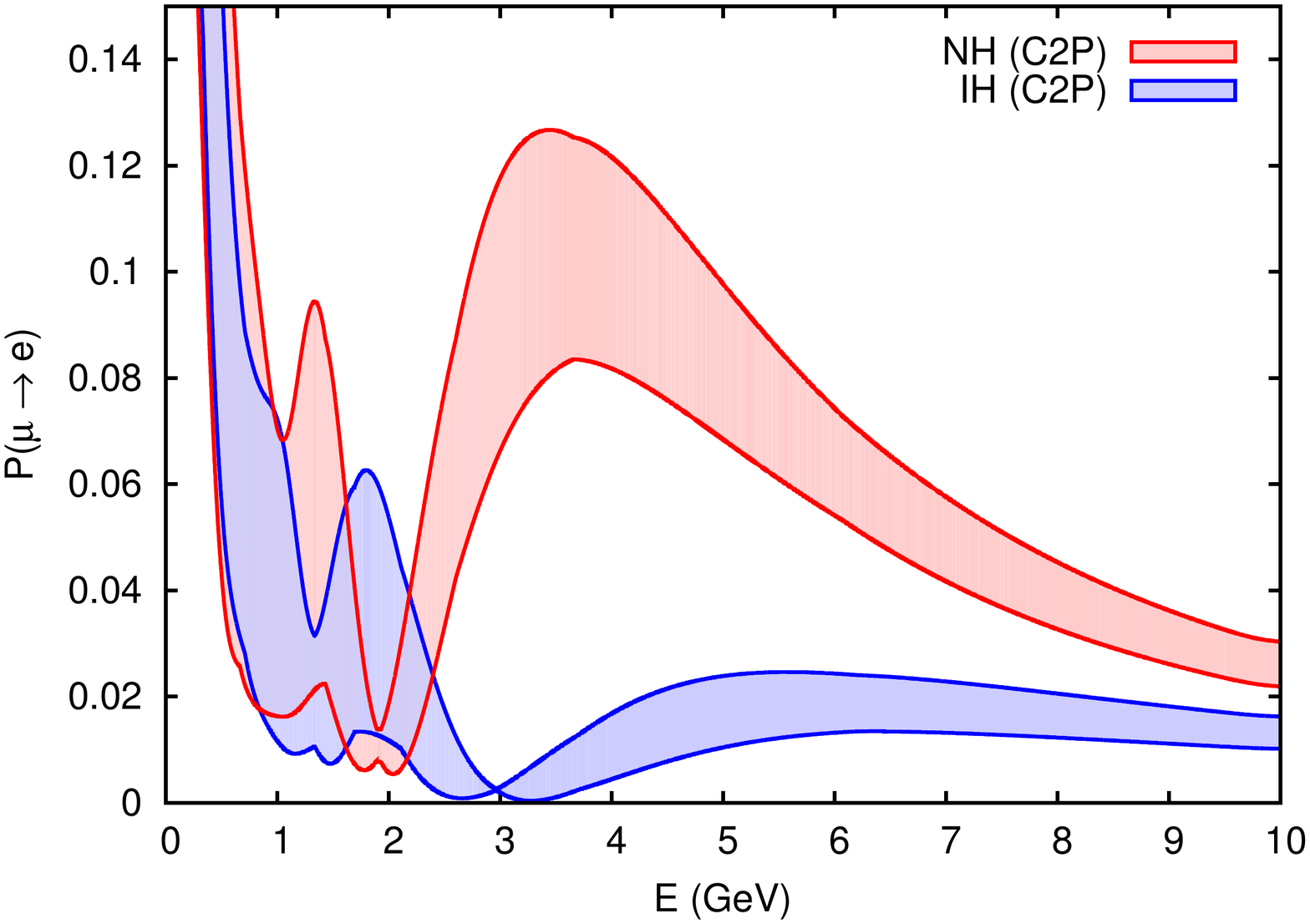}\\
a) Protvino beam & b) CERN beam
\end{tabular}
\caption{$\numu$ to $\nue$ oscillation probability at the Pyh\"{a}salmi site 
for two possible mass orderings: normal (NH) and inverted (IH) hierarchy. The width of 
each band shows the variation in probability induced by the parameter $\dcp$.
\label{fig:prob_NH_IH}}
\end{center}
\end{figure}

For sufficiently long baselines, the mass hierarchy can be determined thanks to the effect on the
oscillation probabilities of the passage
through the Earth. Matter effects modify the propagation, 
with an enhancement of the $\nue$ appearance probability for the neutrinos and
a suppression for the antineutrinos for a normal hierarchy.  If the mass
ordering is inverted, the
opposite happens. By comparing the oscillation probabilities for neutrinos and
antineutrinos, it is possible to deduce MH. The magnitude of this effect
increases as a function of baseline distance, which can be seen in
\reffig{fig:prob_NH_IH}. The longer baseline from CERN to
Pyh\"asalmi leads to a more pronounced separation for the two mass orderings,
especially around the first oscillation maximum. In the following we will
not show the impact of adding P2P to determine the mass hierarchy
since this will be determined in few years of running by C2P alone~\cite{Agarwalla::2014tca}.

\subsection{Benefits of the dual baseline configuration}

It is known since a long time that the inference of the oscillation parameters in long-baseline experiments is
 complicated by the problem of degeneracies~\cite{Barger:2001yr}.
In vacuum three such
degeneracies can be identified: \emph{i)} the intrinsic degeneracy under which
$\theta_{13}$ and $\dcp$ have clone values; \emph{ii)} the sign
degeneracy, as in vacuum it is possible to change the sign of $\Delta m^2_{31}$
and $\dcp$ to $\pi - \dcp$ without affecting the probabilities. This
degeneracy is broken by matter effects. Sufficiently long baselines, for
which matter effects are more significant, are therefore preferred; \emph{iii)}
the octant of $\theta_{23}$, if this angle is not maximal. At present, the
octant is not known.

The known problem of the degeneracies can be mitigated by having access to a
wide band oscillation spectrum, where the information at different
energies acts in a complementary fashion to resolve any ambiguities. We have studied 
and reported the optimisation
of such a wide-band beam elsewhere\cite{Agarwalla::2014tca}. This strategy is
advanced further with the combination of P2P and C2P, where information is
collected from two broad spectra focused on overlapping yet different parts of the
$L/E$ spectrum.


\section{Experimental setup\label{sec:protvino}}
\subsection{The Protvino accelerator complex}
\begin{figure}[h]
\begin{center}
\includegraphics[width = 0.60\textwidth]{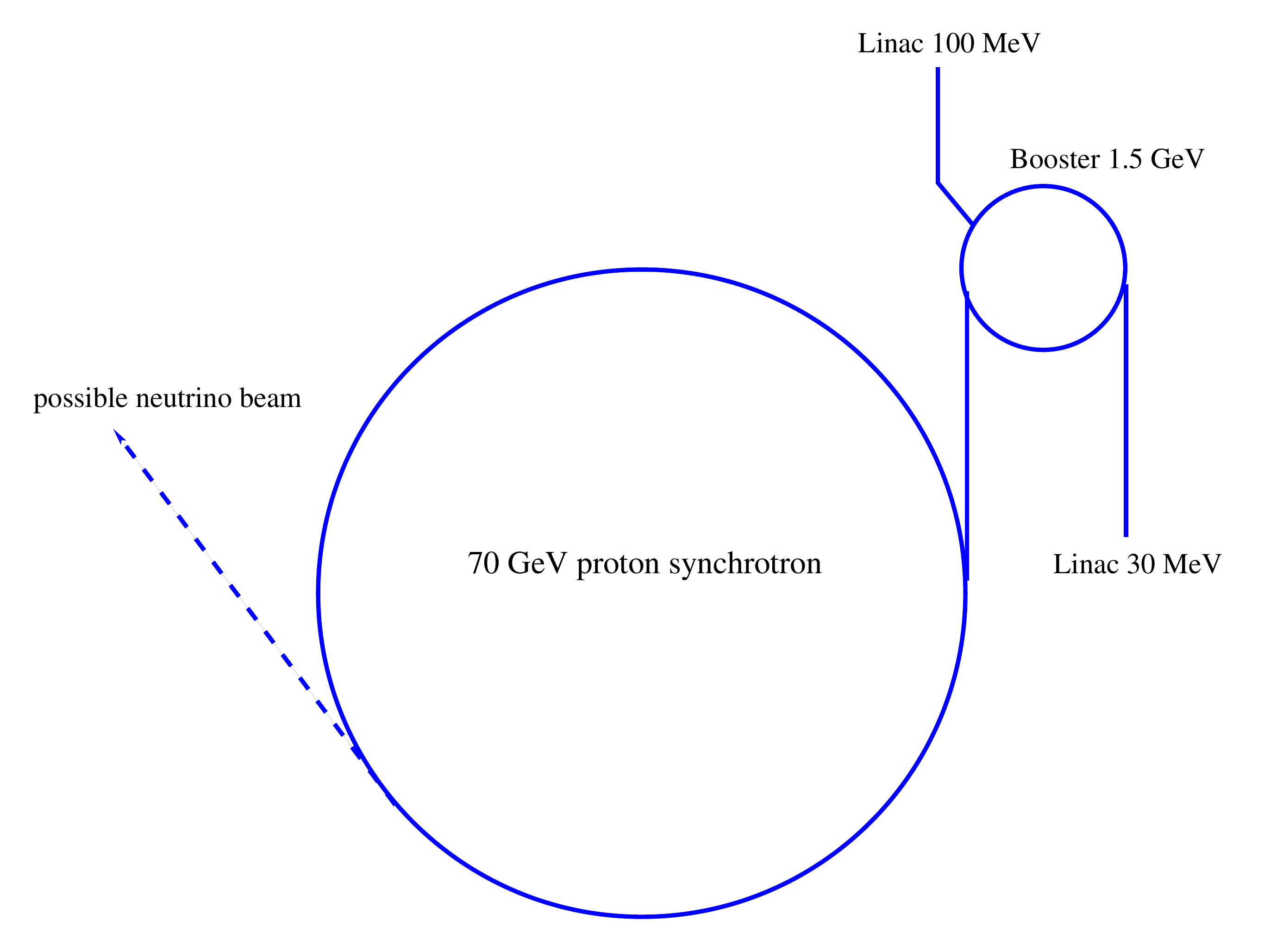}
\caption{Schematic view of the IHEP accelerator complex.}
\label{fig:complex}
\end{center}
\end{figure}

The IHEP accelerator complex consists of four accelerators connected in a 
cascade as  shown in \reffig{fig:complex}.
A 30 MeV linear accelerator serves as an injector for a 1.5 GeV rapid cycling
booster synchrotron, a 100 MeV Alvarez type linear accelerator which serves as
a light-ion or a back-up proton injector feeding the 1.5 GeV synchrotron, and
a 70 GeV proton synchrotron which now operates  at 50 GeV with a beam power of
8-15 kW. The experimental program includes studies of rare kaon decays, hadron
spectroscopy, spin physics and hadron-nuclei interactions.   
A dedicated neutrino beam line has also been constructed and
used for several experiments at U-70, e.g. the IHEP-JINR Neutrino Detector~\cite{Barabash:2002zd}.
An upgrade of a
beam power to 400-450 kW is envisioned in the future as an extension of the
Omega Project~\cite{omega}. We assume in the following that this power
is available for a new neutrino beam line directed towards Pyh\"asalmi.

\subsection{Simulation of the neutrino beamlines}
\label{sec:beam}

\begin{figure}[tbhp]
\begin{center}
\includegraphics[width = 0.99\textwidth]{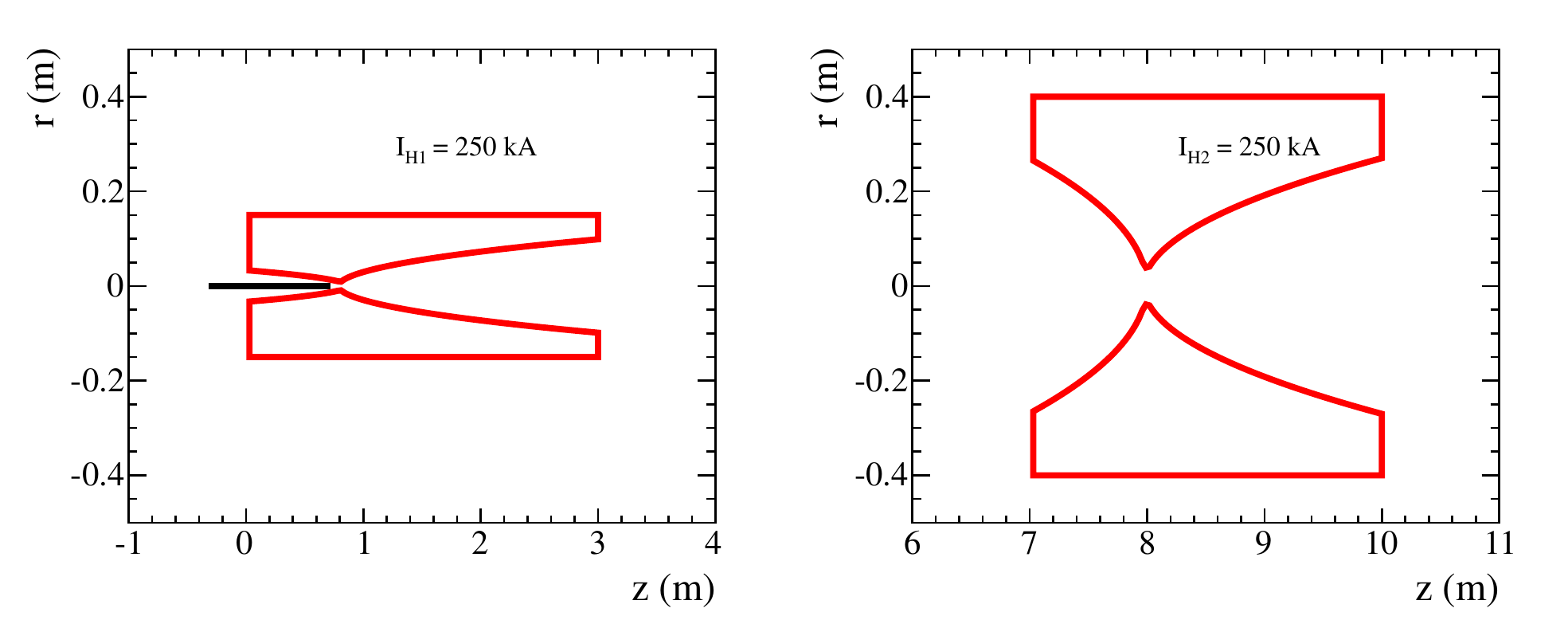}
\caption{Illustration of the target and horn geometry.}
\label{fig:targhorn}
\end{center}
\end{figure}

For this study, we assume a proton beam energy of $70$~GeV and a power of 450~kW
on the neutrino target.  The neutrino beam
simulation for the Protvino site is based on the simulation code developed in the
context of the first phase LAGUNA design study \cite{Longhin:2012ae}. The hadron
production target is modeled as a cylinder, which is $4$~mm in diameter and
$1$~m long and is made from light density ($\rho = 1.85$~g/cm$^3$) graphite.
The focusing system consists of two horns with parabolic inner conductors. The
shape of the inner conductors as well as the distance between the horns are
taken to be the same as in \cite{Stahl:2012exa}. A two-dimensional layout of the
horn-target system is shown in \reffig{fig:targhorn}.

To optimize the energy spectrum of the neutrino beam we varied the
longitudinal position of the target and the currents in the horns. The values of these
parameters were randomly drawn from uniform distributions with realistic
limits. The merit of each configuration was then evaluated based on a
sensitivity to CPV. The optimal value for the target position was found to be
$-0.3$~m upstream of the first horn, while the optimal horn current was
determined to be $250$~kA~\cite{Stahl:2012exa}. 
The radius and the length of the decay tunnel were fixed to
$1.5$~m and $400$~m, respectively. 

\begin{figure}[tp]
\begin{center}
\subfigure{\includegraphics[clip, trim=5 4 6 3, width = 0.44\textwidth]{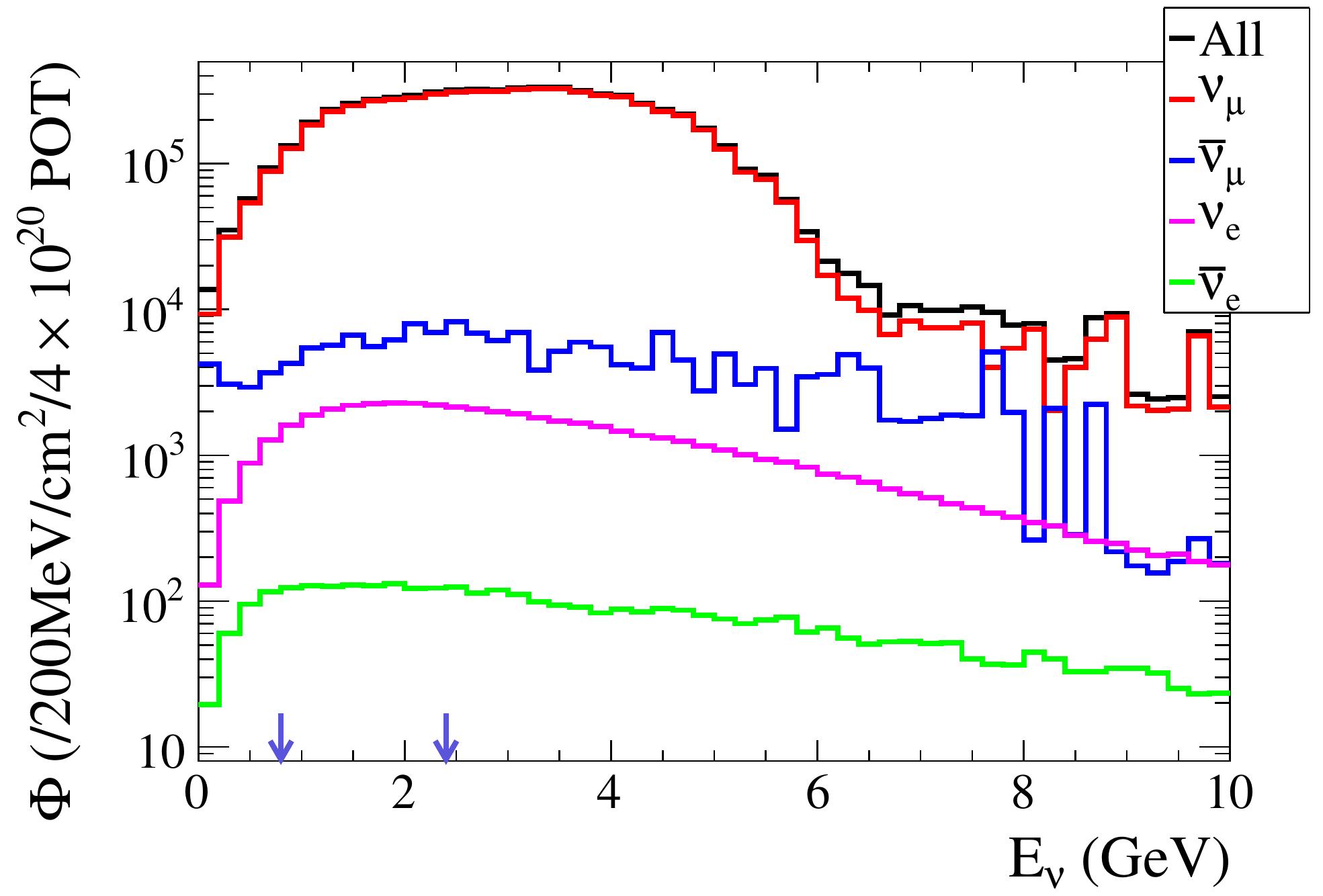}}
\hspace{0.1cm}
\subfigure{\includegraphics[clip, trim=5 4 5 3, width = 0.44\textwidth]{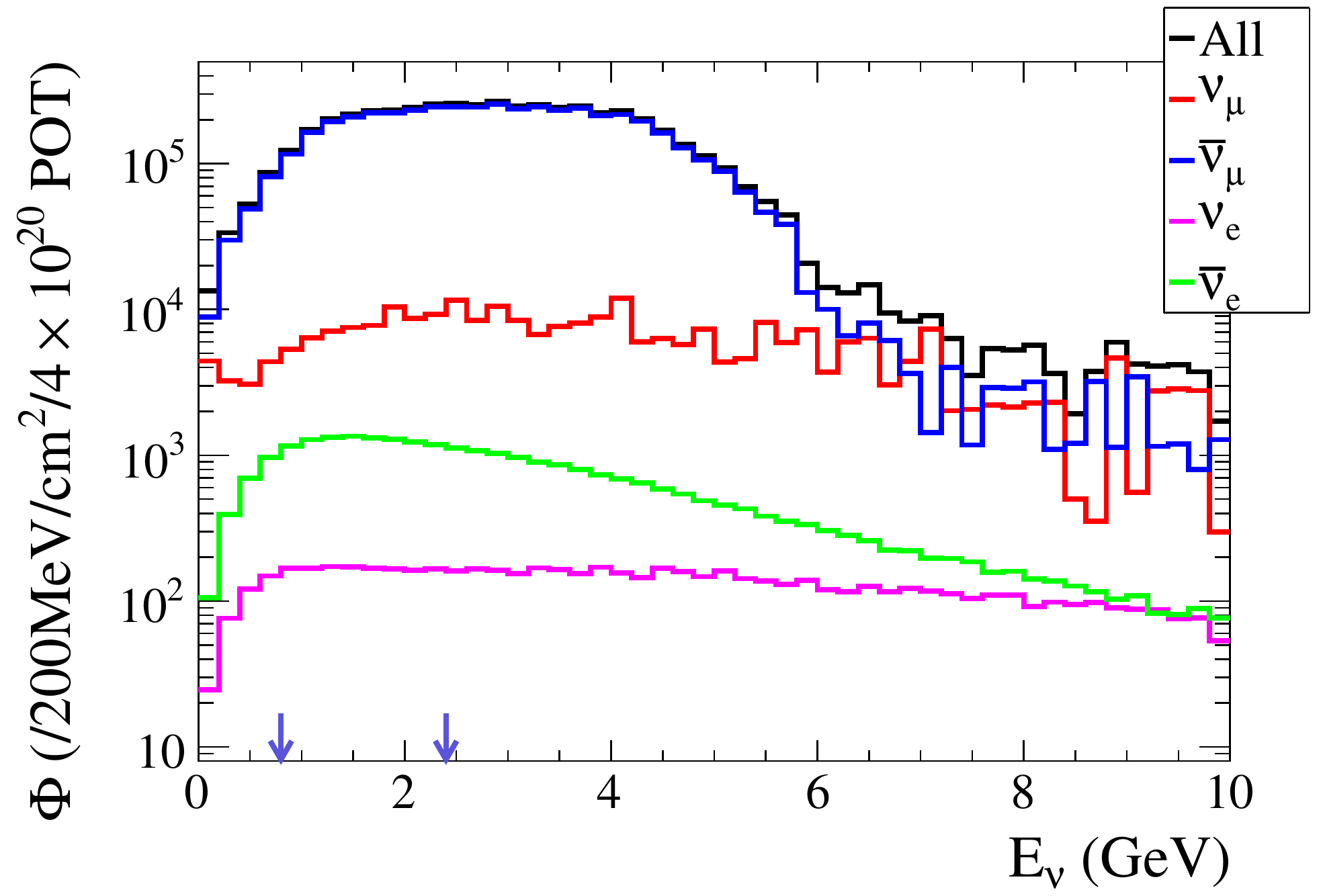}}
\caption{Neutrino (left) and antineutrino (right) fluxes for Protvino-to-Pyh\"asalmi (P2P). The arrows indicate the energy of the first (arrow on the right) and second (arrow on the left) oscillation maxima.}
\label{fig:fluxp2p}
\end{center}
\end{figure}

\begin{figure}[tp]
\begin{center}
\subfigure{\includegraphics[clip, trim=5 4 6 3, width = 0.44\textwidth]{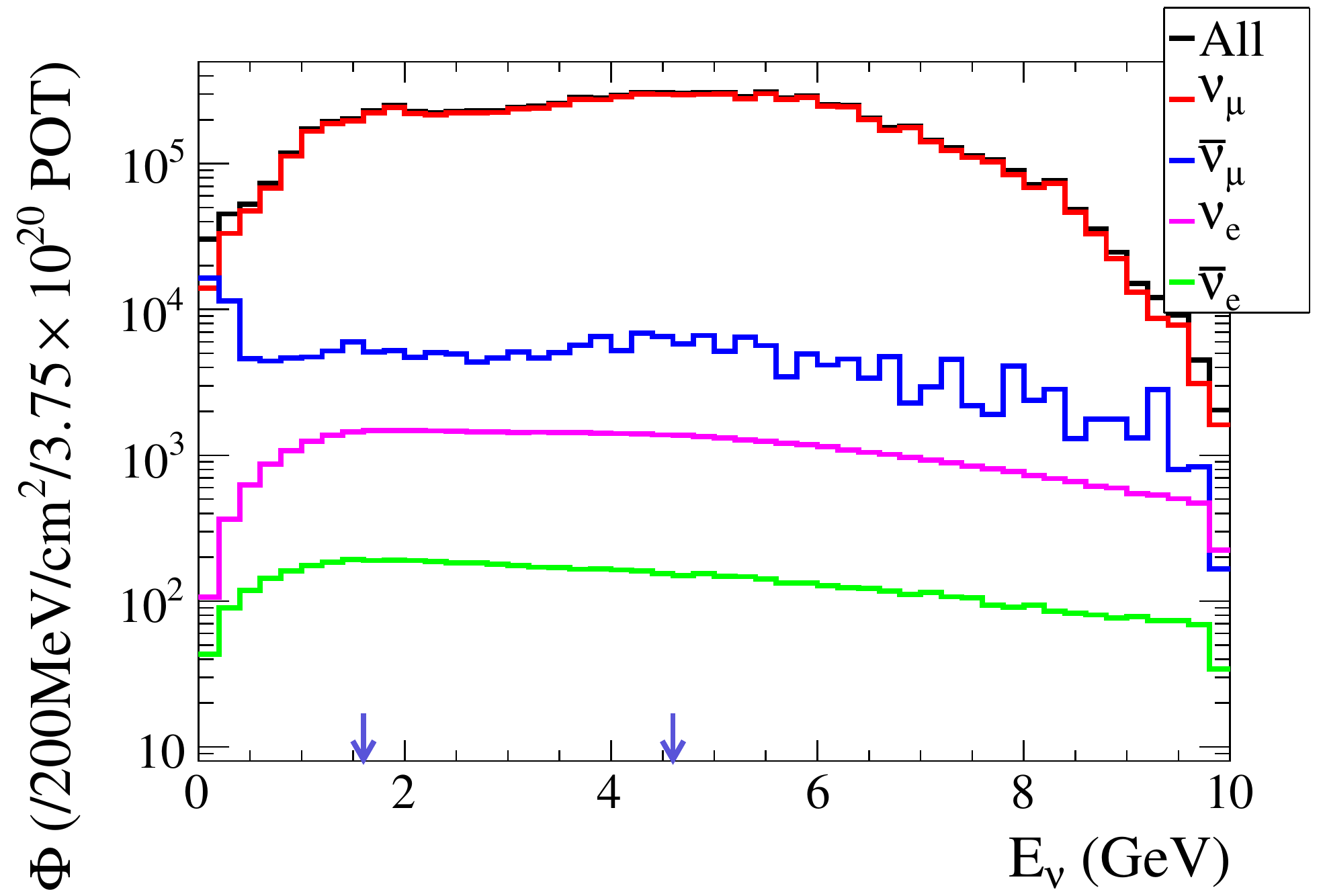}}
\hspace{0.1cm}
\subfigure{\includegraphics[clip, trim=5 4 5 3, width = 0.44\textwidth]{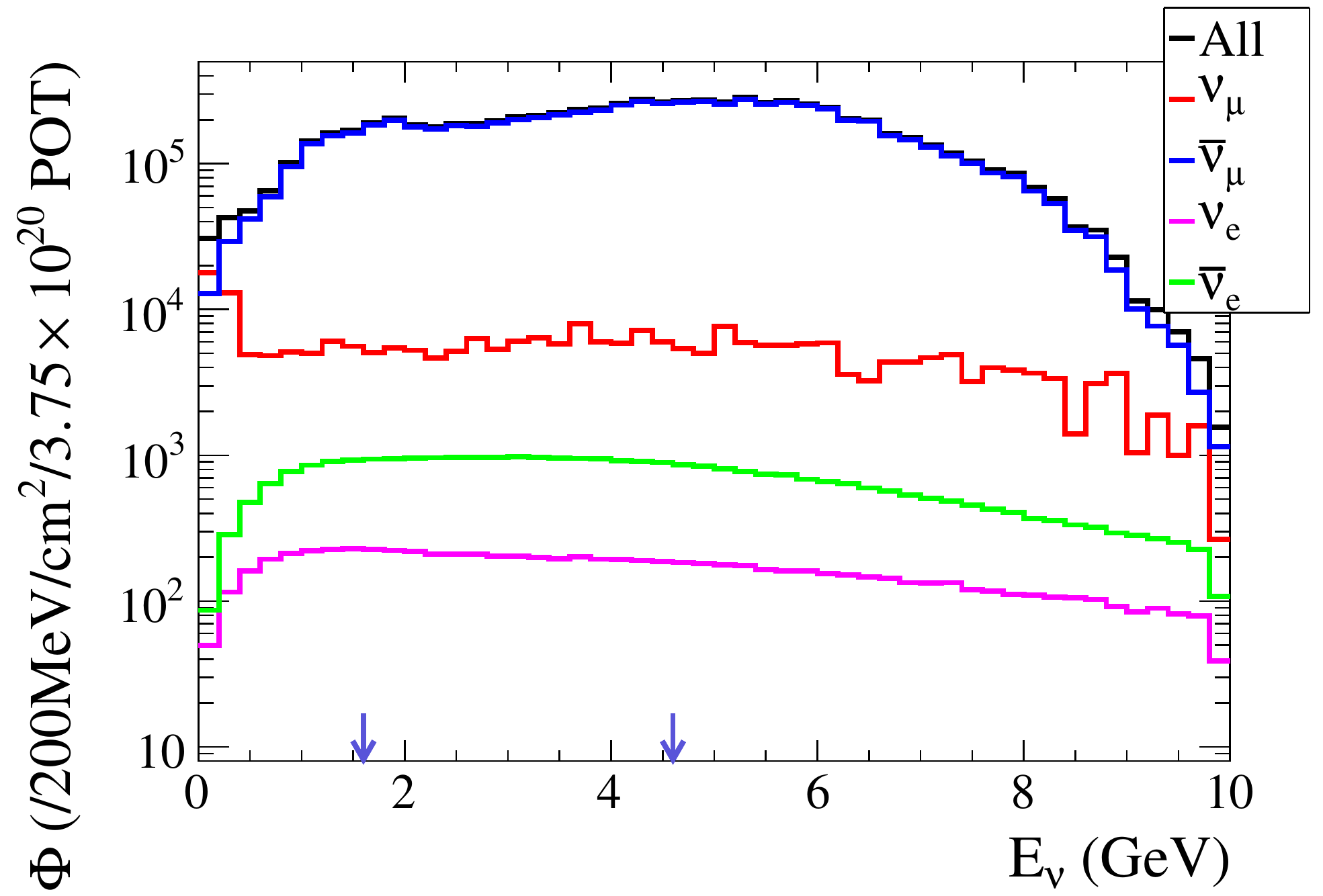}}
\caption{Neutrino (left) and antineutrino (right) fluxes for CERN-to-Pyh\"asalmi (C2P). The arrows indicate the energy of the first (arrow on the right) and second (arrow on the left) oscillation maxima.}
\label{fig:fluxc2p}
\end{center}
\end{figure}

The composition of the non-oscillated fluxes for the P2P and C2P baselines are shown
in \reffig{fig:fluxp2p} and \reffig{fig:fluxc2p}. These fluxes are normalized to $4\times 10^{20}$ protons
on target. With a $450$~kW operation this number corresponds to about one year
of accelerator running with fast extraction, where a year is assumed to
correspond to $10^7$ s. For the simulation of C2P, we have followed
the new optimisation of the SPS fluxes described in
\cite{Agarwalla::2014tca}. The so-called GLB optimisation has been used as
default and the low energy option has been used for cross-check. 
A summary of the proton beam parameters used in this paper for C2P and P2P
are shown in \reftab{tabl:C2PP2Pparam}

\begin{table}[h]
\centering
\caption{\label{tabl:C2PP2Pparam} Assumed parameters of the primary protons
 for the C2P and the P2P beams.}
\begin{tabular} { l  c  c } \hline
{\bf Parameter} & {\bf C2P beam } & {\bf P2P beam} \\ \hline
$E_\text{beam}$ (GeV) & 400 & 70 \\
$I_\text{beam}$ (ppp)  & $7\times10^{13}$ & $2.2\times10^{14}$ \\
Cycle length (s)	&  6 & 5 \\
$P_\text{beam}$ (kW)	& 750 & 450  \\
POT$_\text{year}$ ($10^{20}$) & $1.0\div1.4$ & 4 \\  \hline
\end{tabular}

\end{table}


\section{CP violation determination in the dual baseline configuration}
\label{sec:anameth}
In this section we focus on the benefit of the dual baseline
configuration for the discovery of CP violation in the leptonic
sector. We do not show results for the determination of the MH since
the C2P setup alone already provide a guaranteed measurement
($>5\sigma$) of the mass hierarchy within few years~\cite{Stahl:2012exa,Agarwalla::2013kaa,Agarwalla::2014tca}.

In order to compare setups, we have performed simulations of the P2P and C2P cases both individually
and in combination.  The primary channel of interest is the appearance channel,
$\nu_\mu \to \nu_e (\overline{\nu}_\mu \to \overline{\nu}_e)$, as it is
particularly sensitive to sign($\Delta m^2_{31}$) and $\dcp$.  
We have also included the disappearance channel as it plays an important role in
the precise determination of the value of $\Delta m^2_{31}$ and $\theta_{23}$,
and indirectly on the precision with which one can determine
$\dcp$. The true value of
$\theta_{23}$ has a significant influence on the sensitivity to
$\dcp$~\cite{Agarwalla::2013kaa}. The solar parameters ($\Delta m^2_{31}$  and
$\theta_{13}$) are fixed in the fit since their impact on the results is negligible.

We consider the following background contributions to the signal
e-like events:
\begin{itemize}
\item{Intrinsic $\nu_e$ contamination in the beam (intrinsic $\nu_e$),}
\item{Electron events from $\nu_\tau$ charged current interaction with subsequent leptonic $\tau$ decay ($\nu_\tau \rightarrow e$ contamination),}
\item{Neutral current $\nu_\mu$ events with $\pi^0$ production (NC~$\pi^0$),}
\item{Mis-identified muons from $\nu_\mu$ CC interactions (mis-id $\nu_\mu$).}
\end{itemize}
A detailed description of neutrino event simulations and selection efficiency can
be found in~\cite{Stahl:2012exa}. 

\subsection{Experimental observables}
\label{subsect:recovar}

\begin{figure}[tb]
\centering
\subfigure[~All e-like]{
	\includegraphics[clip, trim = 10 0 30 20, width=0.32\textwidth]{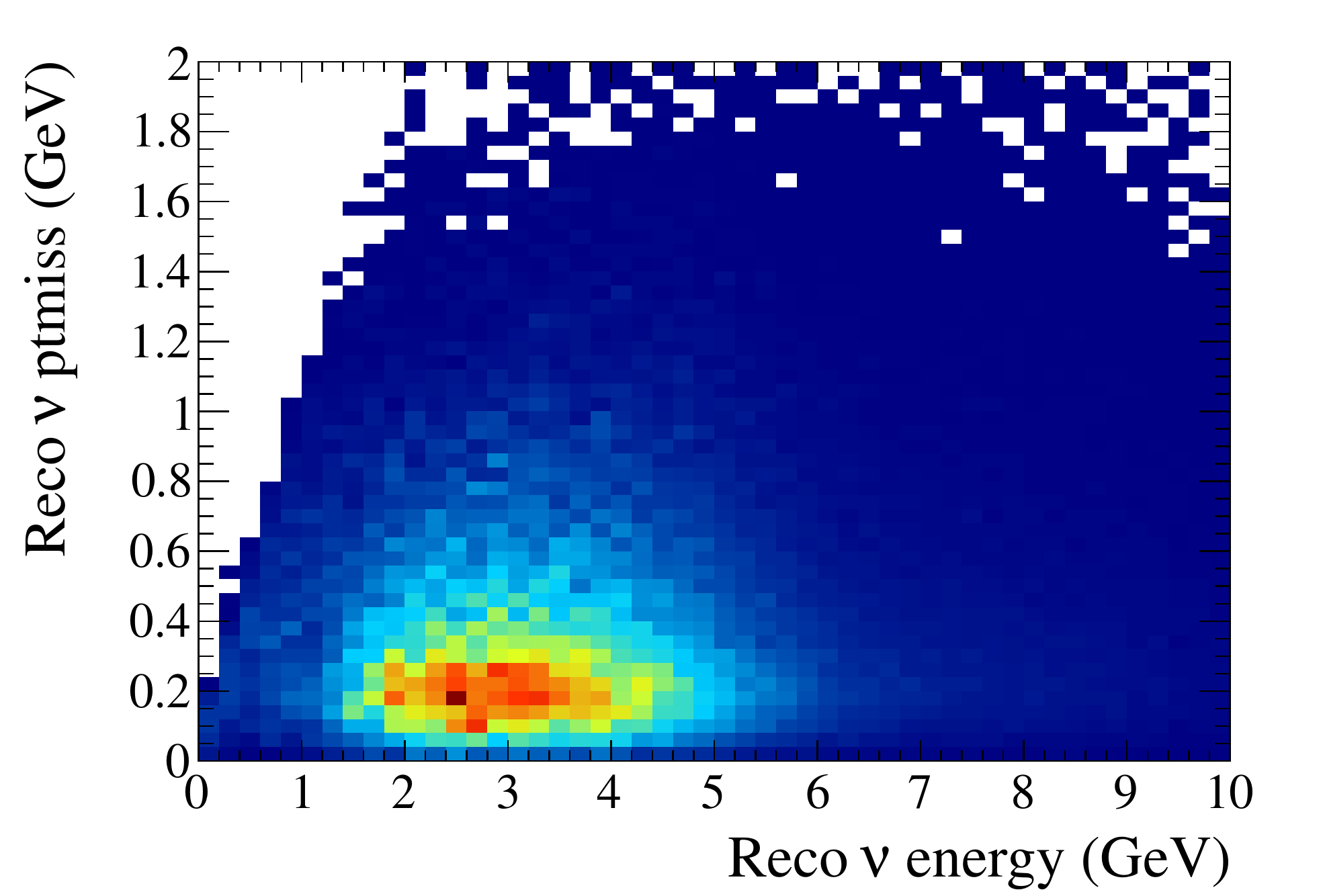}}
\subfigure[~Signal $\nu_e$]{
	\includegraphics[clip, trim = 10 0 30 20, width=0.32\textwidth]{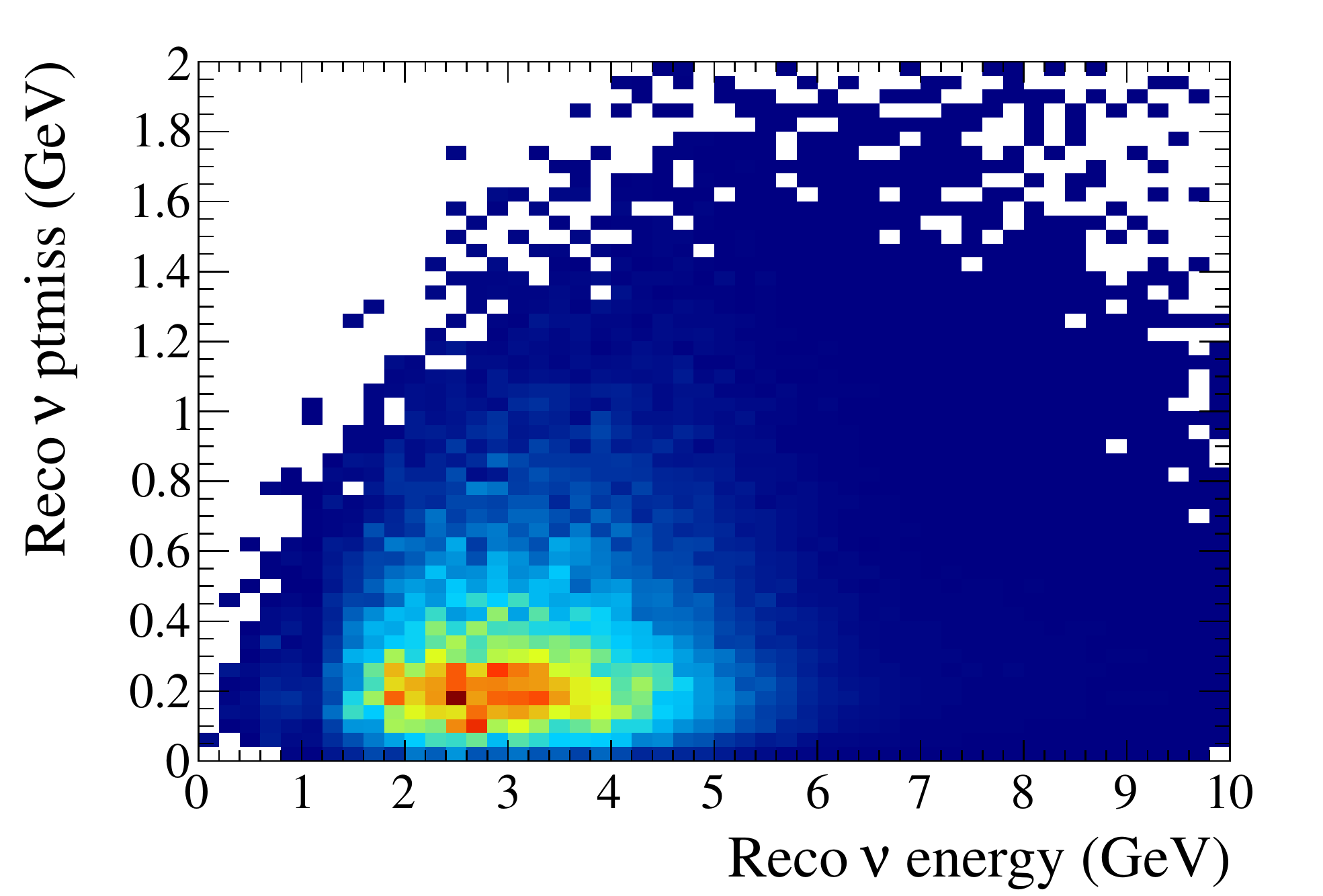}}
\subfigure[~Intrinsic $\nu_e$]{
	\includegraphics[clip, trim = 10 0 30 20, width=0.32\textwidth]{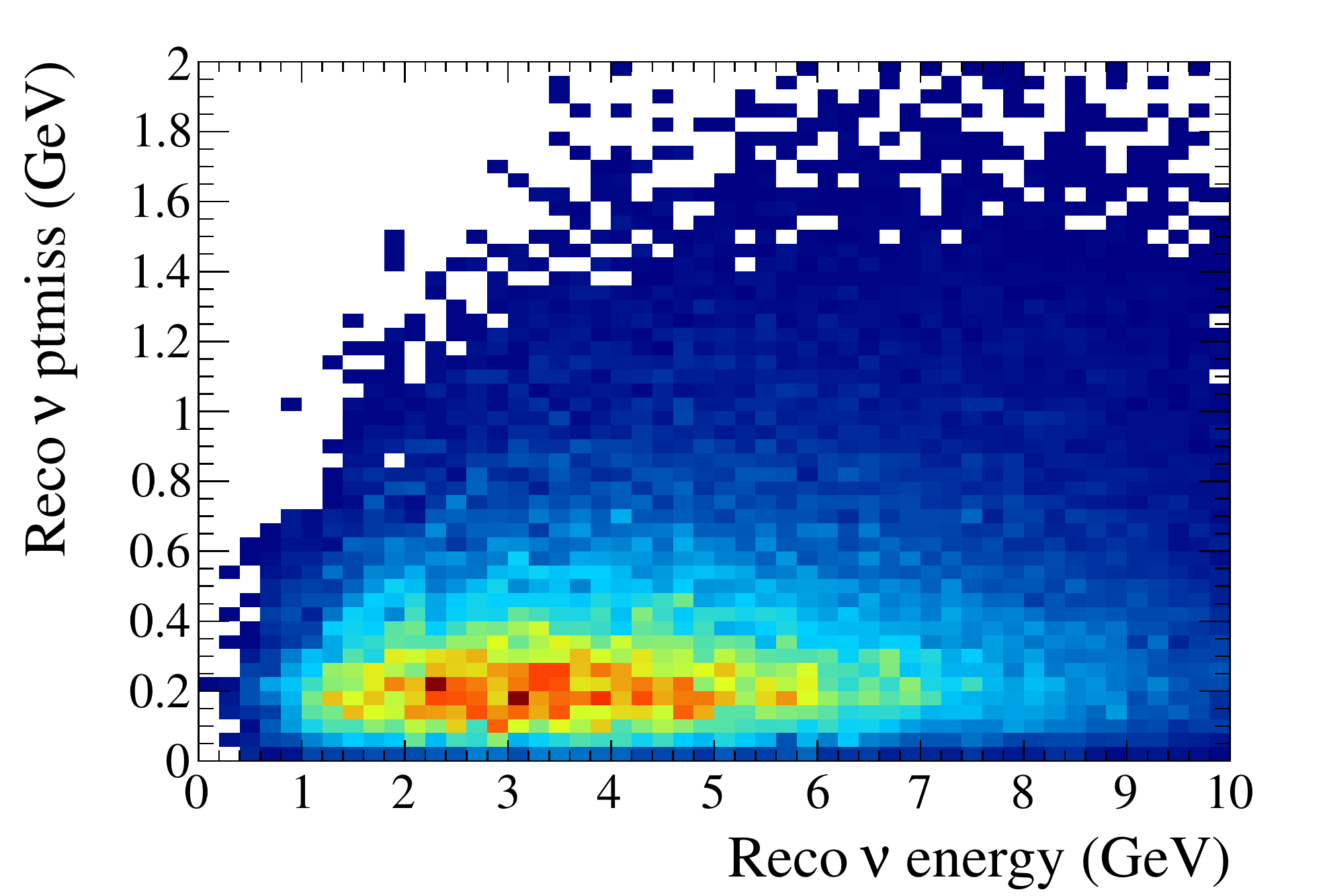}}

\subfigure[~NC~$\pi^0$]{
	\includegraphics[clip, trim = 10 0 30 20, width=0.32\textwidth]{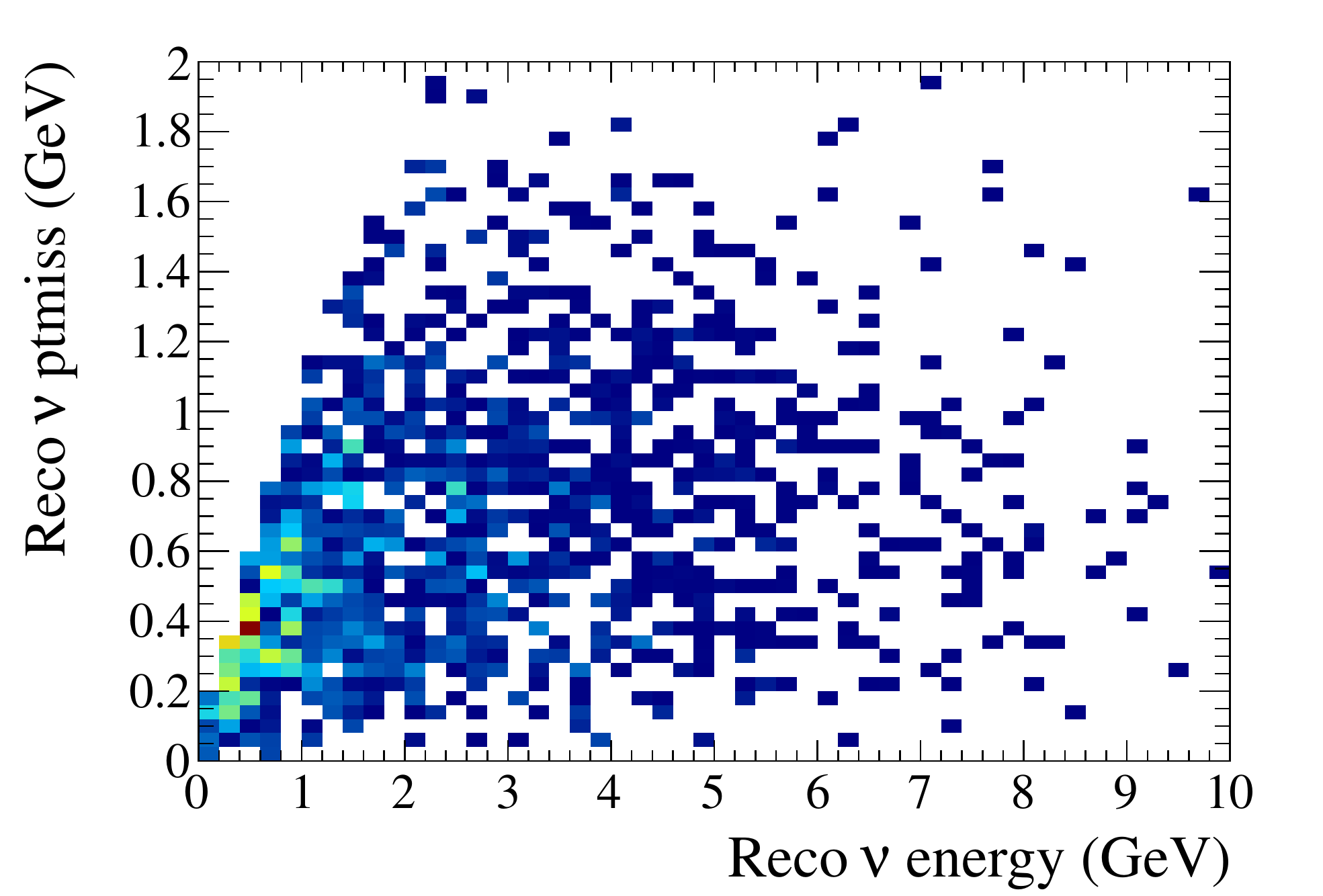}}
\subfigure[~$\nu_\tau \rightarrow e$ contamination]{
	\includegraphics[clip, trim = 10 0 30 20, width=0.32\textwidth]{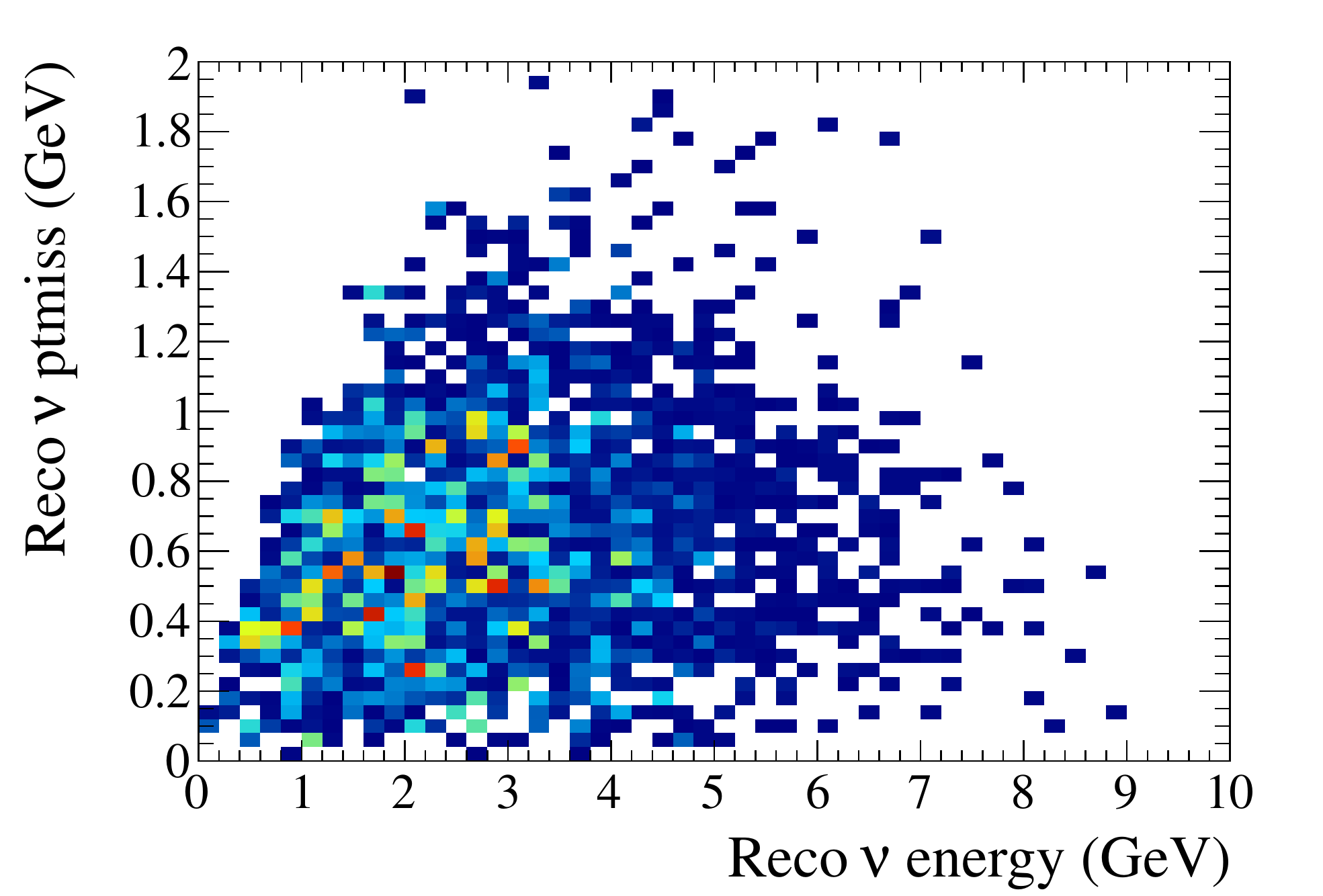}}
\subfigure[~Mis-id $\nu_\mu$]{
	\includegraphics[clip, trim = 10 0 30 20, width=0.32\textwidth]{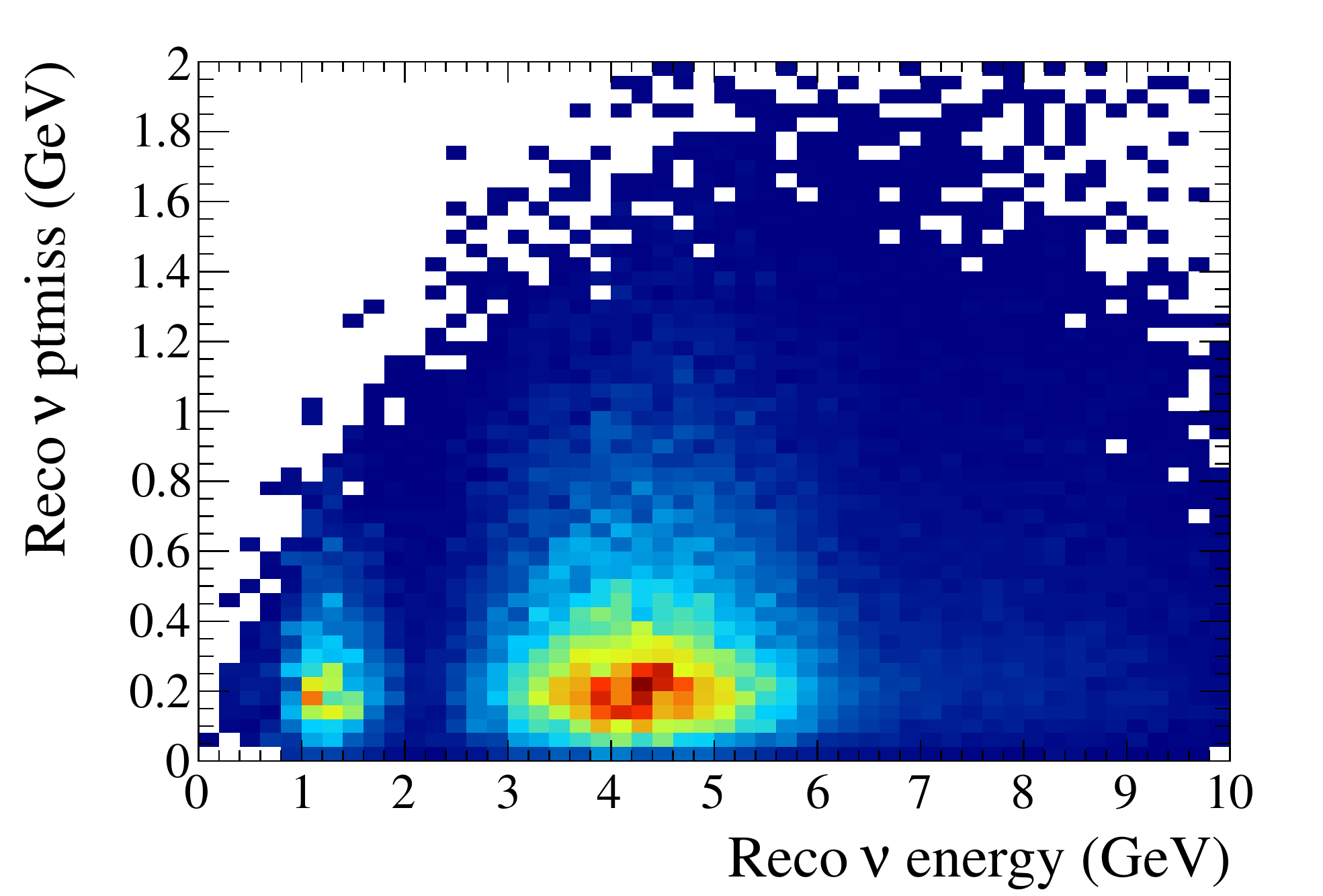}}
\caption{Example event distributions in the $E_{\nu}^{rec}-p_T^{miss}$
  phase-space for various channels contributing to the P2P e-like
  sample. $\dcp=0$ and NH hierarchy are assumed.}
\label{fig:elikedist}
\end{figure}

As experimental observables we use reconstructed neutrino energy, $E_{\nu}^\text{rec}$, and missing
momentum in the transverse plane, defined by the incoming neutrino
beam direction, $p_T^\text{miss}$, of each e-like event to construct
bi-dimensional distributions used to discriminate signal from background. Examples of such distributions are shown in
\reffig{fig:elikedist} for a value of $\delta_{CP} = 0$ and the case of normal
mass hierarchy for the P2P case. As can be seen in the figure, the shape of the signal and
background contributions in the $E_{\nu}^\text{rec}-p_T^\text{miss}$
phase-space differ. 
In particular, NC $\pi^0$ interactions are characterized by low $E_{\nu}^{rec}$ values, while events originating from $\nu_\tau$ CC interactions tend to have larger $p_T^{miss}$ than the $\nu_e$ CC events because of the two neutrinos in the final state.
This allows a better signal-background discrimination than
if one were to use $E_\nu^\text{rec}$ information only. 
In the case of the $\mu$-like events only the reconstructed neutrino
energy is used since the background components in this
sample are smaller.

\subsection{Fit and analysis method}
\label{subsec:anameth}

The key ingredient to be provided to the fit are the priors for the oscillation
parameters $\mathbf{o}$ and for the detector systematic uncertainties $\mathbf{f}$. 
For both sets of uncertainties we have used the same parameters
described for the LBNO CPV reach using CERN beams. 
The assumed values of the oscillation parameters  $\mathbf{o}$ are given in
\reftab{tab:sumoscparam}. The parameters whose values are not labeled as
``exact'' are allowed to vary in the fit. 
\begin{table}[h]
\centering
\begin{tabular}{c|c|c}
\hline
\hline
Parameter                  & Central value                    & Uncertainty \\
\hline
$\Delta m^2_{21}$      & $7.45\times 10^{-5}\mbox{ eV}^2$ & exact  \\
$|\Delta m^2_{32} |$   & $2.5\times 10^{-3}\mbox{ eV}^2$ & 3.75\% \\
$\sin^2{\theta_{12}}$  &  $0.31$                        & exact  \\
$\sin^2{2\theta_{13}}$ & $0.09$                         & 3\%   \\
$\sin^2{\theta_{23}}$  &  $0.45$                         & 5\%   \\
\hline
\end{tabular}
\caption{Assumed values of the oscillation parameters.}
\label{tab:sumoscparam}
\end{table}

\reftab{tab:sumcorrparam} summarizes
the assumptions on the Earth density and systematic uncertainties on the
expected number of events for the different components contributing to
the electron-like sample. Due to the difference in the baselines, the
average matter densitiy encountered by each neutrino beam may not be
the same. The Earth density for C2P have been extensively studied in
the context of LAGUNA-LBNO and it was found to be 3.2~g/cm$^3$. For
the Protvino beam we have checked that the CPV sensitivity does not
depend on the Earth density if this parameter is varied between 3.2
and 2.8~g/cm$^3$. In the combination the value of the Earth density is
fixed to 3.2~g/cm$^3$ for both the baselines. 

\begin{table}[ht]
\centering
\begin{tabular}{c|c|c|c}
\hline
\hline
Name & Central value & Uncertainty & Correlation ($\rho_i$) \\
\hline
Earth density (g/cm$^3$)  & 3.2  & 4\%  & $[0, 1]$ \\
Signal event normalization $f_\text{sig}$  & 1.0  & 3\%  & $[0, 1]$ \\
Beam $\nu_e$ normalization $f_{\nu_e}$  & 1.0  & 5\%  & $[0, 1]$ \\
Tau event normalization $f_{\nu_\tau}$  & 1.0  & 20\%  & $[0, 1]$ \\
$\nu$ NC and $\nu_\mu$ CC background $f_\text{NC}$  & 1.0 & 10\%  & $[0, 1]$ \\
\hline
\end{tabular}
\caption{Assumptions on the matter density and the uncertainties on the event normalizations for two beams. These parameters can be treated as uncorrelated or correlated between the two beams by changing the value of $\rho_i$.}
\label{tab:sumcorrparam}
\end{table}

In the fit we minimize the following $\chi^2$ with respect to the
oscillation and systematic parameters:

\begin{equation}
\chi^2 = \chi^2_\text{appear} + \chi^2_\text{disa} + \chi^2_\text{syst}.
\label{eq:totchi2}
\end{equation}

The $\chi^2_\text{appear}$ is the term corresponding to the electron-like sample. It is given by the sum over the beamline $i$ (C2P or P2P),
the horns polarity
$j$ (positive or negative) and the bin $k$ in the $E_{\nu}^{rec}-p_T^{miss}$ phase-space:

\begin{equation}
\chi^2_\text{appear} = 2\sum_\text{i}\sum_{j}\sum_\text{k} R^{e\text{-like}}(i,j,k),
\label{eq:chi2appear}
\end{equation}

For each $E_\nu^\text{rec}-p_T^\text{miss}$ bin, $ R^{e\text{-like}}$
is given by:

\begin{equation}
\begin{split}
R^{e\text{-like}} = & n_{e}(E_\nu^{rec},p_T^{miss}; \mathbf{o}_{test}, \mathbf{f}_{test}) - n_{e}(E_\nu^{rec},p_T^{miss}; \mathbf{o}_{true}, \mathbf{f}_{true}) \\
& + n_{e}(E_\nu^{rec},p_T^{miss}; \mathbf{o}_{true}, \mathbf{f}_{true})\ln{\frac{n_{e}(E_\nu^{rec},p_T^{miss}; \mathbf{o}_{true}, \mathbf{f}_{true})}{n_{e}(E_\nu^{rec},p_T^{miss}; \mathbf{o}_{test}, \mathbf{f}_{test})}},
\end{split}
\label{eq:elikeR}
\end{equation}
\noindent where the subscript \textit{true} (\textit{test}) refers to the true
(test) values of the $\mathbf{o}$ and $\mathbf{f}$ parameters. The
\textit{true} parameters are those chosen by Nature, while
\textit{test} refer to the parameter at which we compute the
likelihood with respect to the \textit{true} value. The number of the e-like events in a given $E_\nu^{rec}-p_T^{miss}$ bin is determined according to:

\begin{equation}
\begin{split}
n_{e}(E_\nu^{rec},p_T^{miss}; \mathbf{o}, \mathbf{f}) &= f_{sig}n_{e-sig}(E_\nu^{rec},p_T^{miss};\mathbf{o}) \\ &+ f_{\nu_e}n_{\nu_e}(E_\nu^{rec},p_T^{miss};\mathbf{o}) + f_{\nu_\tau} n_{e,\nu_\tau}(E_\nu^{rec},p_T^{miss};\mathbf{o}) \\ &+ f_{NC}(n_{NC\pi^0}(E_\nu^{rec},p_T^{miss};\mathbf{o})+n_{mis-\nu_\mu}(E_\nu^{rec},p_T^{miss};\mathbf{o})),
\end{split}
\label{eq:nelike}
\end{equation}
where $n_{e-sig}$, $n_{\nu_e}$, $n_{e,\nu_\tau}$, $n_{NC\pi^0}$, and
$n_{mis-\nu_\mu}$ are the number of events for signal, intrinsic beam
$\nu_e$, electrons from tau decay, neutral current and mis-identified
$\nu_\mu$ respectively.

For the disappearance channel, the information is contained
in the $\chi^{2}_\text{disa}$ term of total $\chi^{2}$ in
\refeq{eq:totchi2}. Similarly to
\refeq{eq:chi2appear} $\chi^{2}_\text{disa}$ is given by:
\begin{equation}
\begin{split}
\chi^2_\text{disa} = 2\sum_\text{i}\sum_{j}\sum_{k} R^{\mu\text{-like}}.
\end{split}
\label{eq:chi2disa}
\end{equation}
\noindent where $i$ refer to the baseline, $j$ to the polarity and $k$ to the
reconstructed energy bin. 
$R^{\mu\text{-like}}$ is calculated in the same way as in \refeq{eq:elikeR}, but using the $\mu$-like event distributions. 
The number of $\mu$-like events in a given bin is the sum of signal
$n_{\mu\text{-sig}}$ and $\tau\rightarrow \mu$ background
$n_{\mu,\nu_\tau}$ contributions:

\begin{equation}
n_{\mu}(E_\nu^{rec}; \mathbf{o}, \mathbf{f}) = f_{sig} n_{\mu-sig}(E_\nu^{rec};\mathbf{o}) + f_{\nu_\tau} n_{\mu,\nu_\tau}(E_\nu^{rec};\mathbf{o}).
\label{eq:nmulike}
\end{equation}

The prior constraints of the oscillation parameters and the
systematics are contained in $\chi^2_\text{syst}$ that has a different
form for correlated and uncorrelated terms. 

When performing the fit of the dual beam experiment it is important to
properly take into account these potential correlations. In our analysis the oscillation
parameters are always treated as fully correlated between the two neutrino
beams, whilst the normalization parameters listed in \reftab{tab:sumcorrparam} can be treated with
varying degree of correlation. 
This is done by introducing two nuisance
parameters and a correlation coefficient ($\rho_i$) for each systematic uncertainty 
listed in \reftab{tab:sumcorrparam}. For
example, to treat the normalization uncertainty on the signal we introduce the
nuisance parameters $f_\text{sig}^\text{C2P}$ and $f_\text{sig}^\text{P2P}$ for
each $\nu$ beam and a correlation coefficient $\rho_{f_\text{sig}}$. These two
parameters would be varied independently in the fit if $\rho_{f_\text{sig}}
= 0$ or with a degree of correlation fixed by the choice of
$\rho_{f_\text{sig}} > 0$. 

The oscillation and the systematic parameters which are fully correlated between the two
beams are constrained through the $\chi^2_{i,\text{syst}}$ term which
has the form

\begin{equation}
\label{eq:chisys_fcorr}
\chi^2_{i,\text{syst}} = \frac{(a_{i} - a_{0,i})^2}{\sigma_{a_i}^2},
\end{equation}
\noindent where $a_{0,i}$ ($a_i$) is the prior (test) value of the
\textit{i\textsuperscript{th}} parameter and $\sigma_{a_i}$ is the
corresponding prior uncertainty.
 
In the case when the \textit{i\textsuperscript{th}} source of
systematic uncertainty is parametrized with two nuisance parameters
$a_{i,1}$ and $a_{i,2}$ and some correlation coefficient $\rho_i$
(e.g., signal normalization), the constraint term has the form:

\begin{equation}
\label{eq:chisys_partcorr}
\begin{split}
\chi^2_{i,\text{syst}} = \frac{1}{1-\rho_i^2} & \left(\frac{(a_{i,1} - a_{0,i})^2}{\sigma_{a_i}^2} + \frac{(a_{i,2} - a_{0,i})^2}{\sigma_{a_i}^2}  - \frac{2\rho_i(a_{i,1} - a_{0,i})(a_{i,2} - a_{0,i})}{\sigma_{a_i}^2}\right),
\end{split}
\end{equation}
where we assume the same prior value $a_{0,i}$ and uncertainty $\sigma_{a_i}$ for both nuisance parameters.
For the following of this paper we assume a full correlation ($\rho=1$)
for both, oscillation priors and normalization uncertainties. 

In order to define the sensitivity of the experiment to CP violation,
we define the test statistics

\begin{equation}
\Delta\chi^{2} = \chi^{2}_{\delta_{CP}} - \chi^{2}_{best},
\label{eq:dchi2cpv_ts}
\end{equation}
where $\chi^{2}_{\delta_{CP}}$ is the minimized $\chi^2$ of
Eq.~\ref{eq:totchi2} at a fixed value of $\delta_{CP}$ (true or test),
while $\chi^{2}_{best}$ is the minimum $\chi^2$ obtained when
$\delta_{CP}$ is allowed to vary over the full range of possible
values.

The significance to observe CP violation is evaluated by computing
$\chi^{2}_{\delta_{CP}}$ in the two CP conserved cases ($\dcp=0$ and
$\dcp=\pi$),  and taking the smallest $\Delta\chi^2$:
\begin{equation}
\label{eq:dchi2cpv}
\Delta \chi^2 = \min{(\Delta\chi^2_0, \Delta\chi^2_\pi)}.
\end{equation}

In the following section, the systematics are
assumed to be fully correlated between the two baselines but we have
checked that the effect on the CPV discovery potential is negligible
if we change the degree of correlation.

\section{Results}
\label{sec:results}

The assumed experimental setup is described in Ref.~\cite{Stahl:2012exa}.
Following the incremental approach, an intitial 20~kton double
phase LAr LEM-TPC (GLACIER~\cite{Rubbia:2004tz,Rubbia:2009md}) is complemented 
with a magnetized muon detector (MIND~\cite{Abe:2007bi,Cervera:2010rz}). 
A second 50~kton detector can be added to reach a total of 70~kton.
The sensitivity to measure CP violation in the dual baseline was
studied for both detector sizes.
In the simulations performed for this paper, the field cage
of the 20~kton detector is approximated with a  cylinder of radius 33~m and height 20~m, corresponding to 
an instrumented volume of  17100~m$^3$ and an active mass of 23.9~kton.
\begin{figure}[h!] 
\begin{center}
\includegraphics[width = 0.495\textwidth]{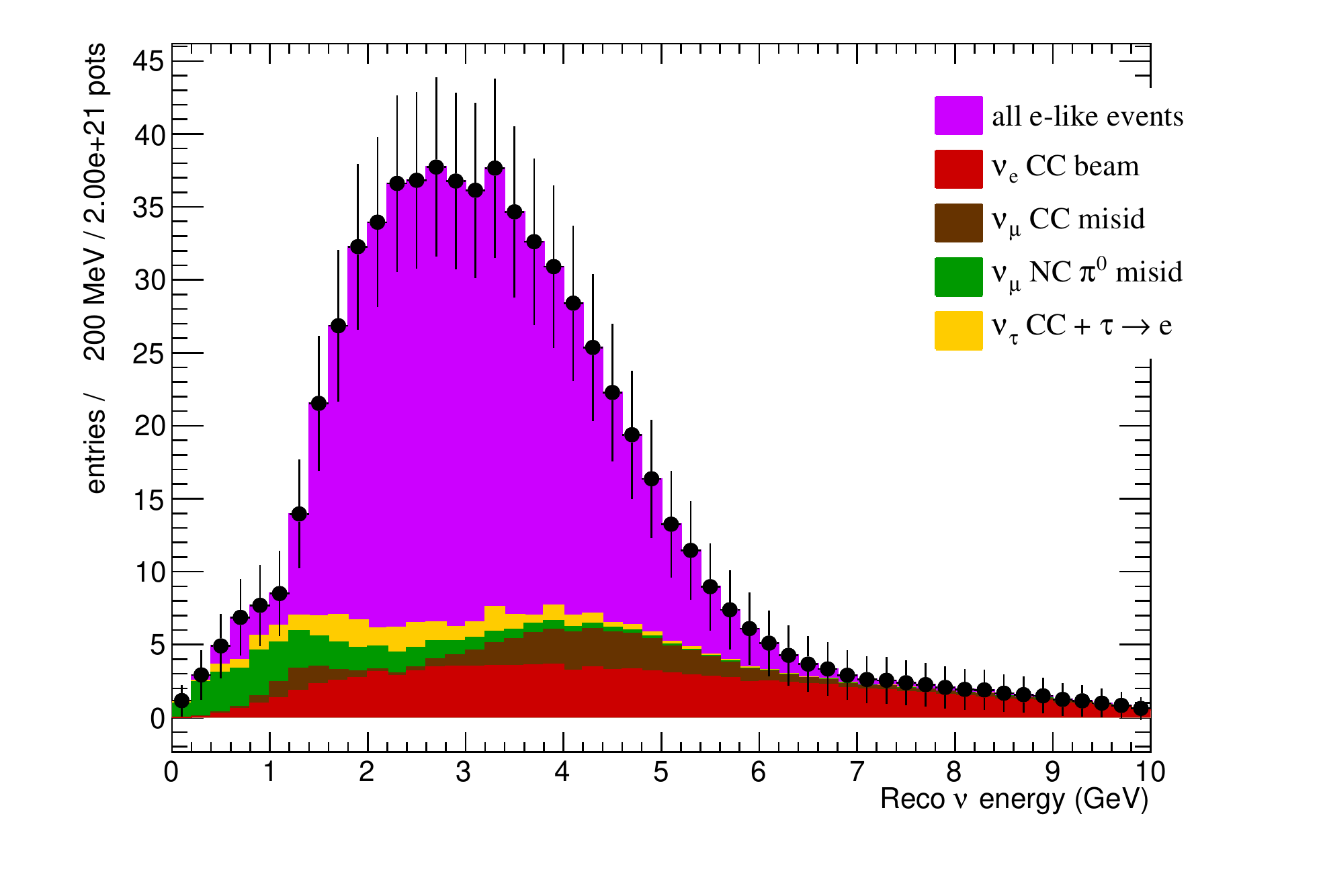}
\includegraphics[width = 0.495\textwidth]{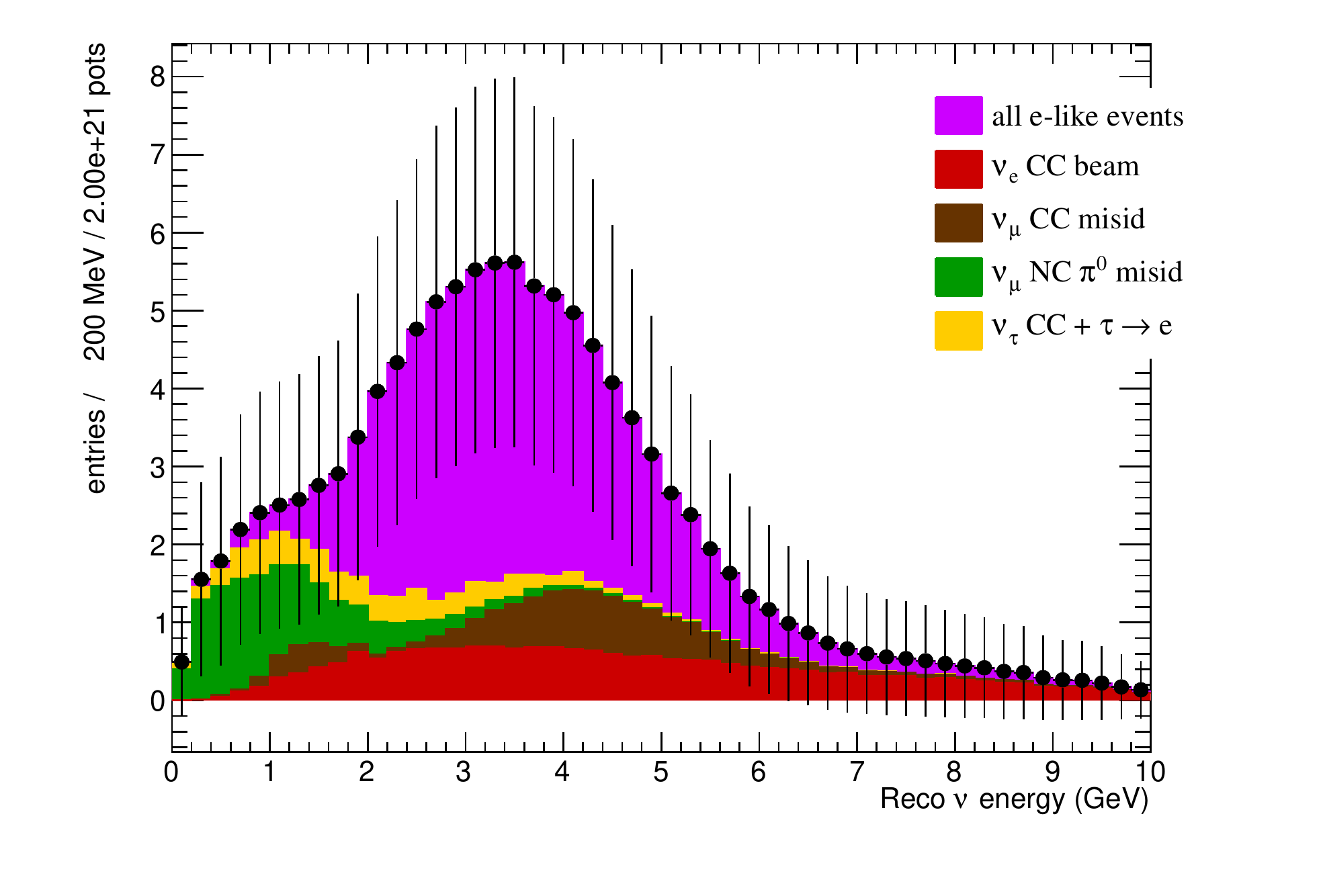}\\
\includegraphics[width = 0.495\textwidth]{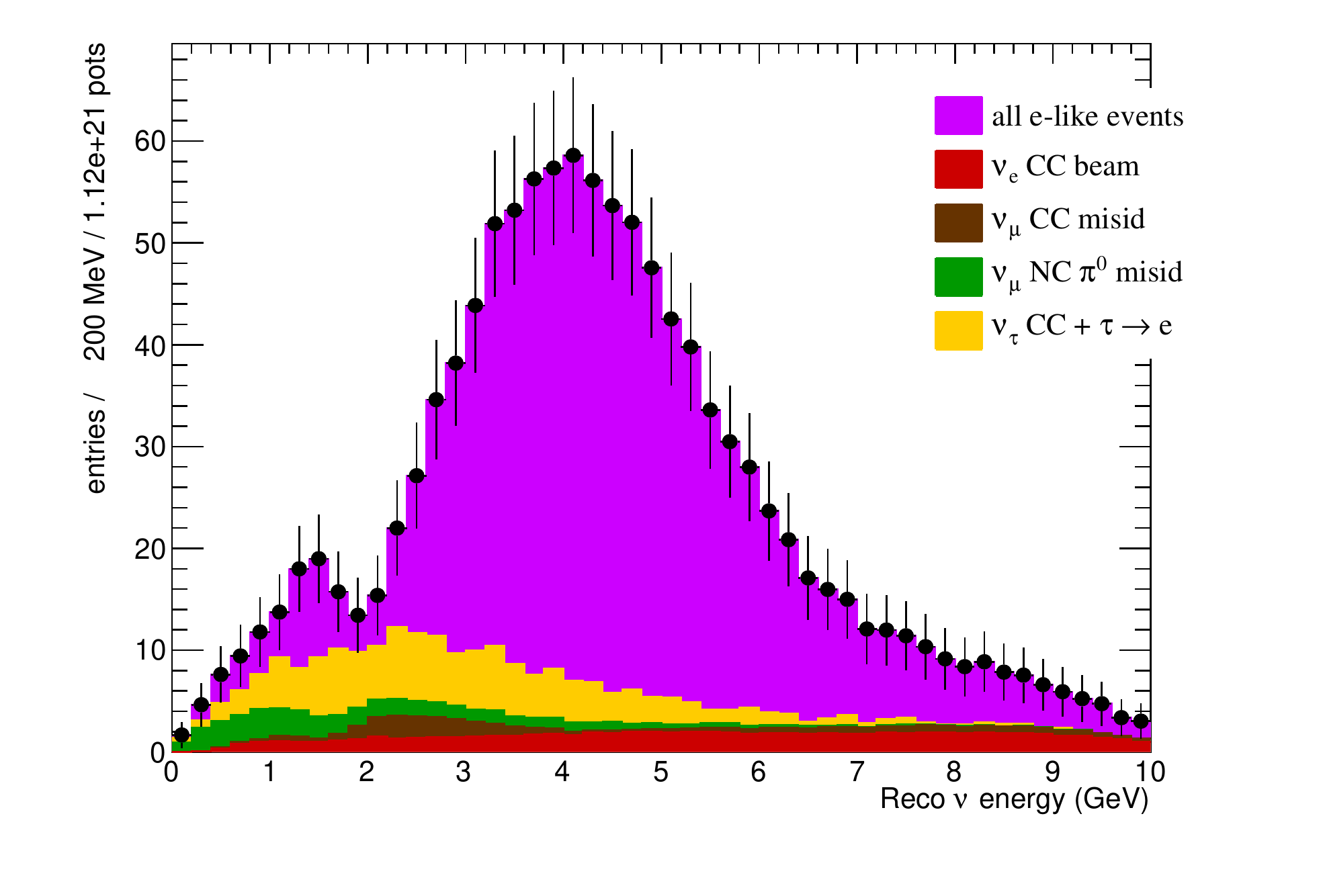}
\includegraphics[width = 0.495\textwidth]{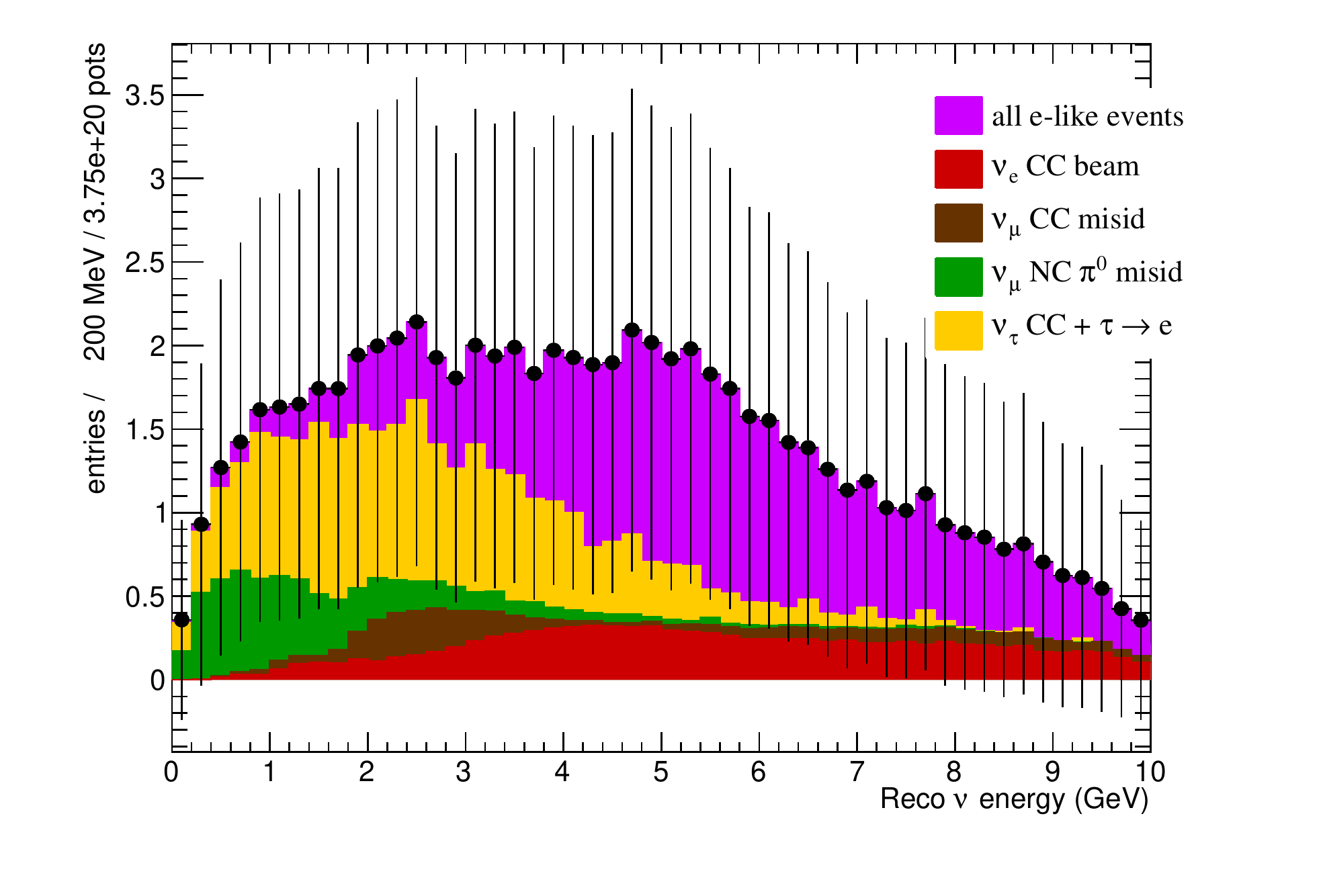}\\
\caption{Reconstructed energy spectra for the Protvino-to-Pyh\"asalmi (P2P) (top) and the 
CERN-to-Pyh\"asalmi (C2P)
  (bottom) in positive horn polarity (left) and negative horn polarity
  (right) for the 20 kton detector. The $\dcp$ is $-\pi/2$ 
  and the hierarchy is normal.\label{fig:events}} \end{center}
\end{figure}

For the P2P beam we assume a total integrated $4\times10^{21}$ POT with a 50\% sharing
of negative and positive polarity while for the C2P beam we take
$1.5\times10^{21}$ POT with a 75-25\% sharing between the negative and
the positive horn polarity using the SPS GLB optimisation
fully described in Ref.~\cite{Agarwalla::2014tca}. In order to illustrate the expected signal, 
the spectrum of  events in the electron-like
sample for $\dcp=-\pi/2$ for P2P and C2P is shown in
\reffig{fig:events}: most of the selected events come from oscillated
neutrinos while the background is due to NC, intrinsic $\nu_e$
and $\nu_{\mu} \rightarrow \nu_{\tau}$ oscillated neutrinos.

\subsection{CP-violation discovery potential for the dual beam facility}
\label{sec:resultsCPV}

The CPV sensitivity the LBNO configuration is provided by the
information contained in the shape of the e-like event
distribution for neutrino and antineutrino modes, 
in particular in the region around the $1^{st}$ and the $2^{nd}$ oscillation
maxima. 

\begin{figure}[h]
\begin{center}
\includegraphics[width = 0.495\textwidth]{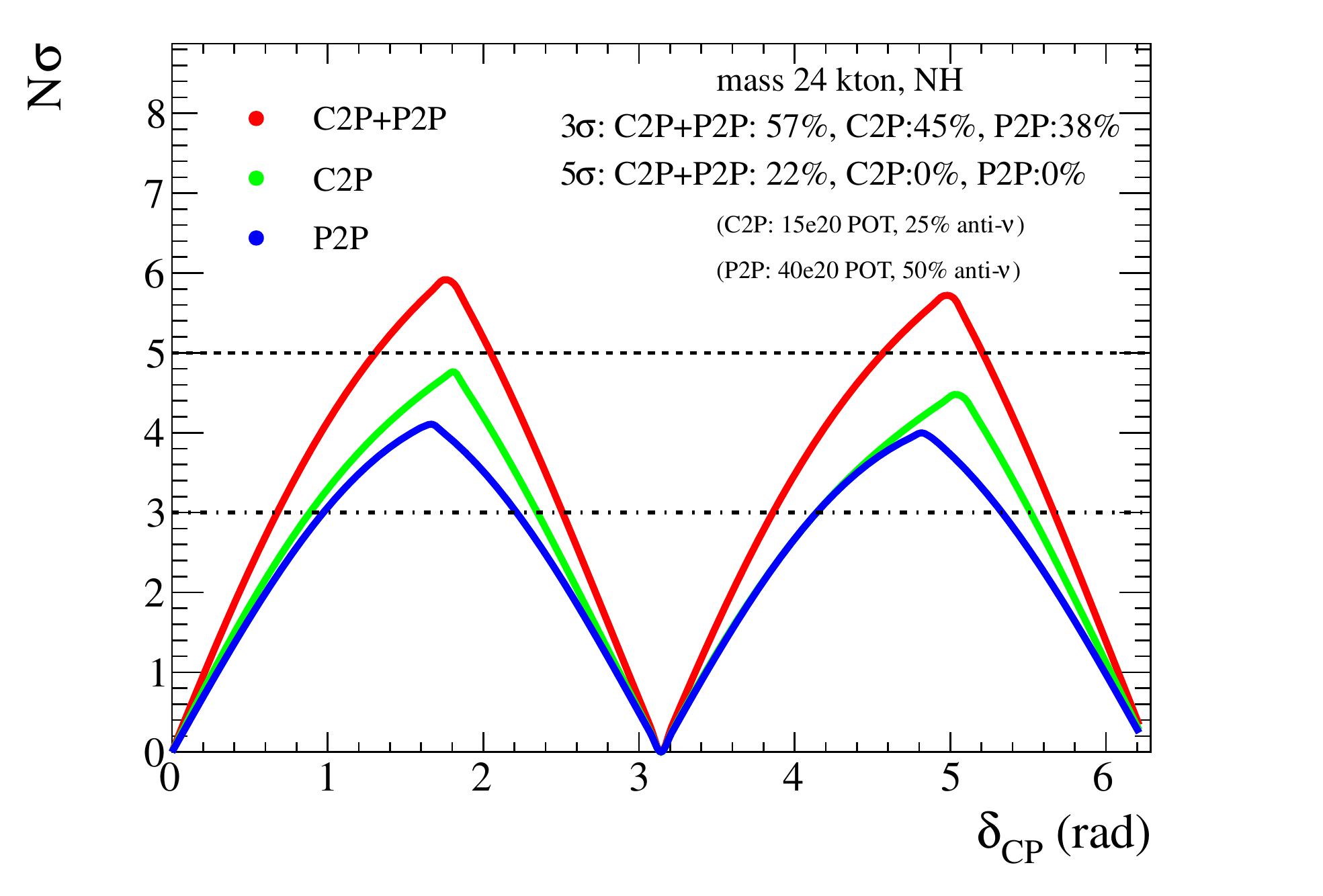}
\includegraphics[width = 0.495\textwidth]{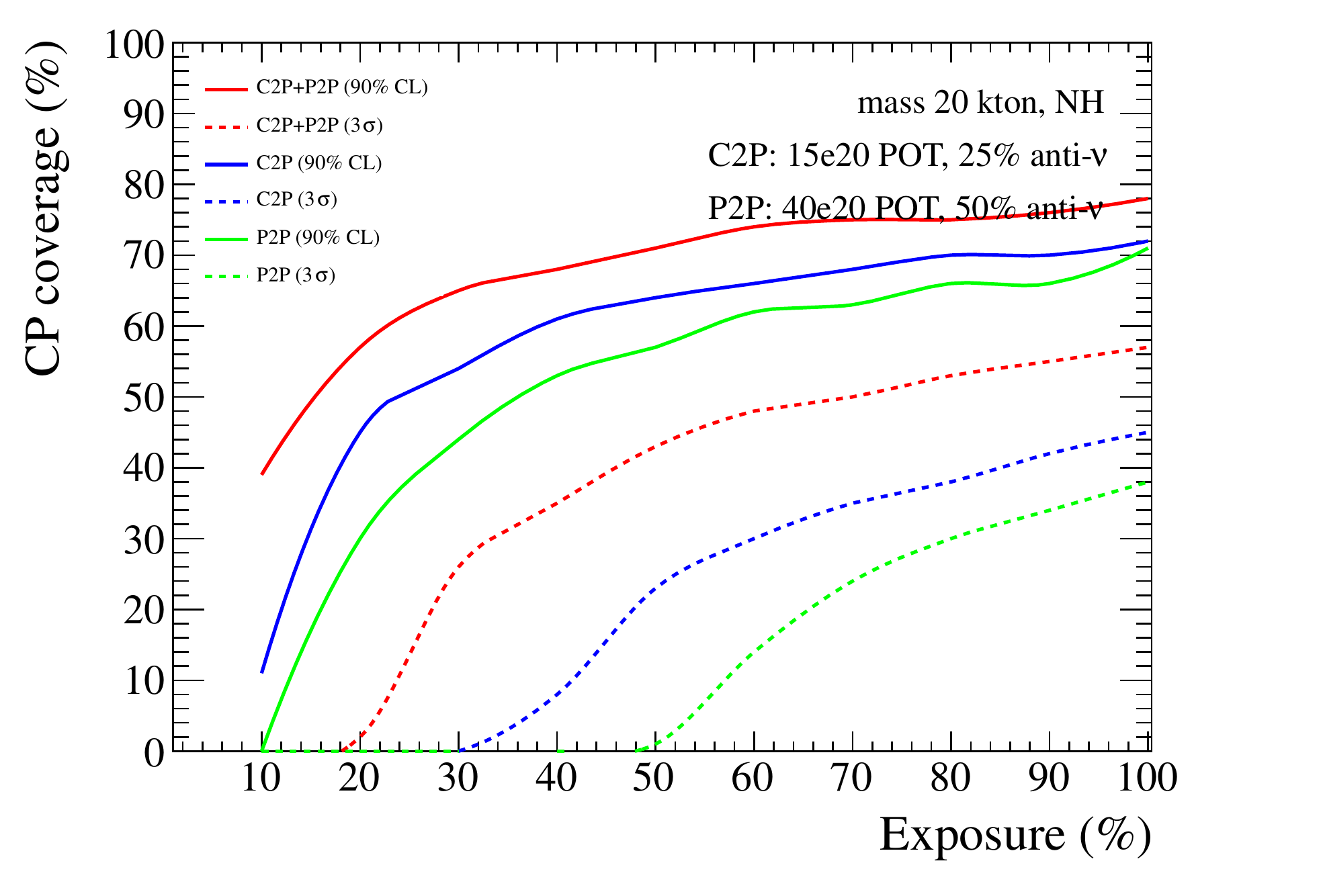}\\
\caption{Left: $\sqrt{\Delta \chi^2}=N\sigma$ as a function of the true value of $\dcp$ for
  CP-violation discovery. Right: Sensitivity
  to CP-violation in terms of the fraction of
  $\dcp$ values for which $\delta=0,\pi$ can be excluded, as a
  function of the total exposure. Both plots are obtained assuming a
  20~kton far detector and NH.}
\label{fig:CPV20}
\end{center}
\end{figure}

The sensitivity as a function of the true value of the $\dcp$ phase is shown in \reffig{fig:CPV20} for C2P
only, P2P only and their combination assuming NH. 
We see that each baseline separately can reach a 3$\sigma$
determination of CP-violation ($\Delta\chi^2>9$) for $\sim40\%$ 
of the values of the $\dcp$ phase. The combination of the two beams,
instead, would allow to establish CP-violation at $3\sigma$ for 
$\sim60\%$ of values of $\dcp$ even with a 20~kton far detector.

\begin{figure}[h]
\begin{center}
\includegraphics[width = 0.49\textwidth]{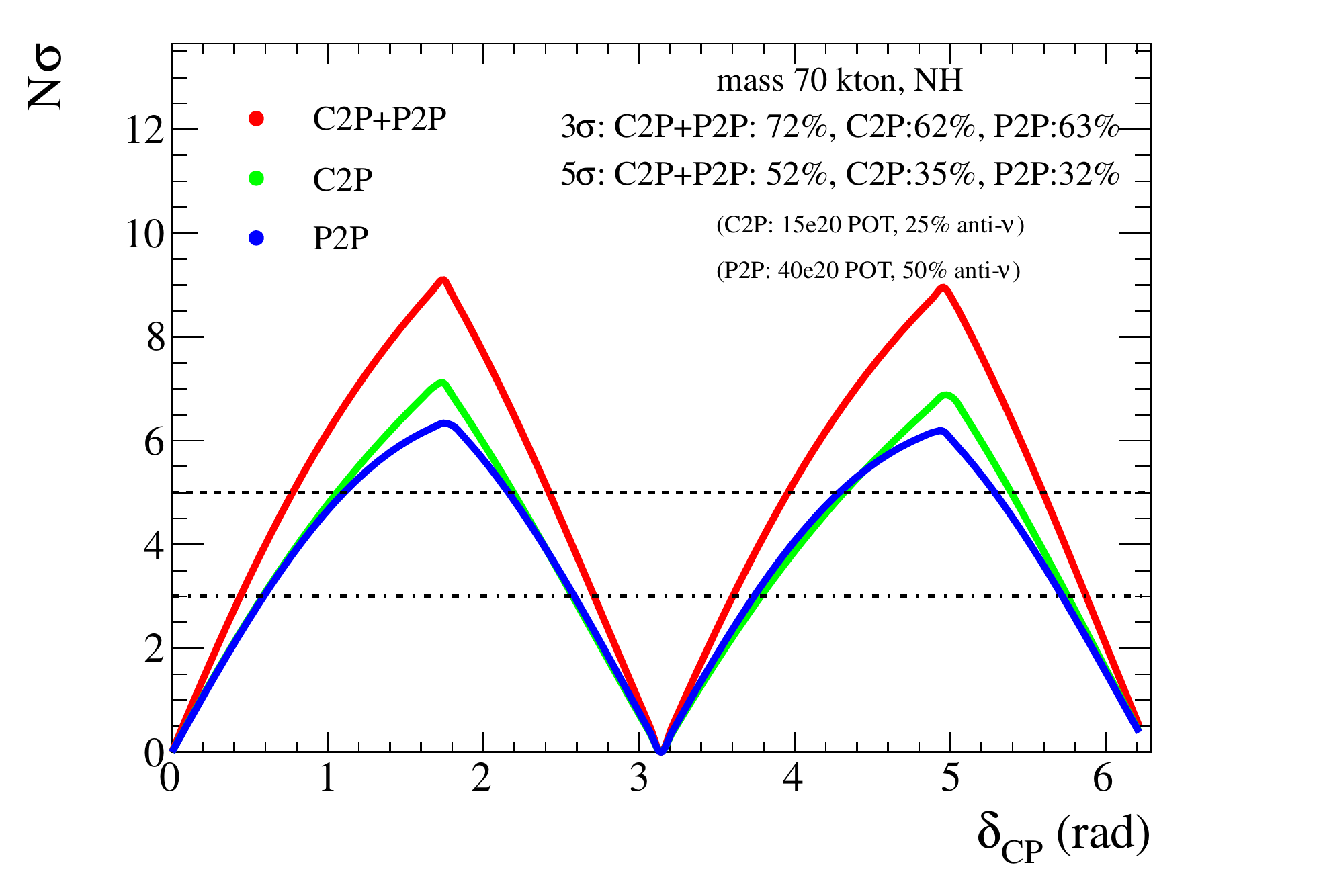}
\includegraphics[width = 0.49\textwidth]{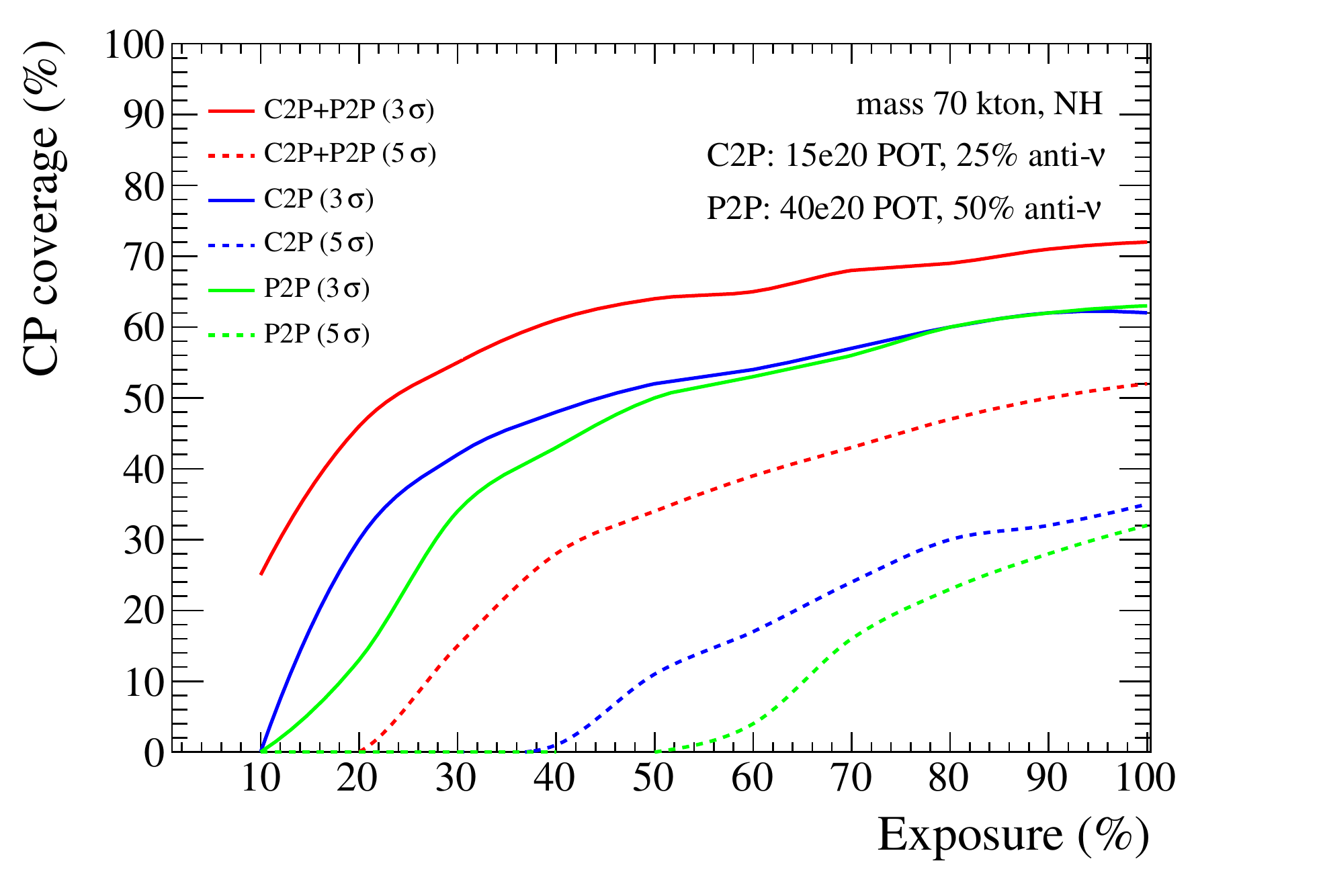}
\caption{Left: $\sqrt{\Delta \chi^2}=N\sigma$ as a function of the true value of $\dcp$ for
  CP-violation discovery. Right: Sensitivity
  to CP-violation in terms of the fraction of
  $\dcp$ values for which $\delta=0,\pi$ can be excluded, as a
  function of the total exposure. Both plots are obtained assuming a
  70~kton far detector and NH.}
\label{fig:CPV70}
\end{center}
\end{figure}

The same sensitivities assuming the same exposure but a larger,
70~kton far detector, are shown in \reffig{fig:CPV70}. A larger detector and the combination of the
two beams would allow to measure CPV at more than $3(5)\sigma$ for the
70(50)\% of the true values of $\dcp$.
Similar results are obtained in the case of inverted hierarchy.


\subsection{Advantages of the dual baseline configuration}
When considering the combination of two beams, there are a number of effects
which lead to better performances. In this section, we will consider
separately the following factors which may improve the sensitivity to $\dcp$:
\begin{itemize} 
\item a dual beam facility allows for a greater number of events, decreasing
the statistical uncertainty;
\item combining beams at the same detector may allow for the cancellation of
systematic uncertainties;
\item both beams have access to different parts of the oscillation parameter
space, and the information that they each provide will be complementary,
increasing the overall sensitivity.
\end{itemize}

The use of two beams significantly increases the total number of
neutrino interactions, without
requiring to accumulate POT from a single accelerator complex for
which it would be necessary to run the experiment for more years. 
Adding P2P will help to reduce the statistical uncertainty of the observed
rates and allow for an improved sensitivity to $\dcp$. 

The additional sensitivity obtained thanks to the combination of C2P and P2P goes beyond the simple additional
statistical power brought by the P2P beam. To establish this, we must
compare the combined beams configuration to a single beam facility with
an increased exposure, chosen to mitigate any statistical advantage. 
This can be obtained by scaling the total POT of the C2P beam to add
the additional statistical power brought by the P2P beam. In our
calculation we assumed a beam power delivered by the Protvino
accelerator with a proton beam energy of 70~GeV and a total exposure of $40\times10^{20}$
POT. The equivalent power of the C2P beam, using a proton beam of 400~GeV is then given by:

\begin{equation} \label{eq:potequiv} N_\text{POT~equiv}^\text{C2P} =
40\times10^{20} \mathrm{POT} \frac{70}{400} = 7 \times 10^{20} \mathrm{POT}.  \end{equation} 

\begin{figure}[h]
\begin{center}
\includegraphics[width=0.6\textwidth]{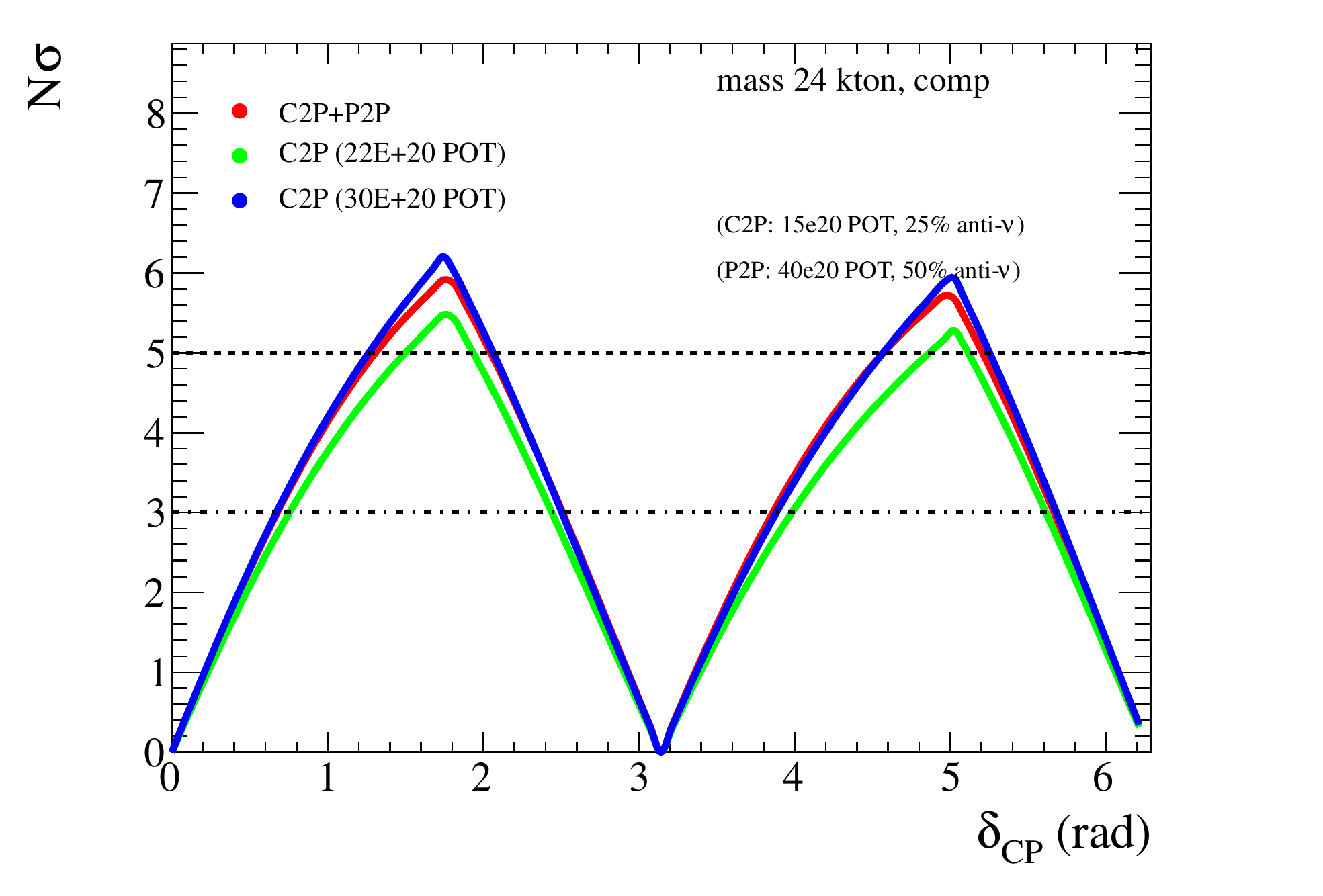}
\caption{Comparison of the CPV sensitivity obtained with two neutrino beam and
a 20 kton far detector to the CPV sensitivity obtained with a single beam and a
larger far detector.} \label{fig:comparecpv} 
\end{center}
\end{figure}

Consequently in the single C2P beam scenario one has to increase the
total number of POT to $22\times10^{20}$ POT.
To remove the statistical advantage of the dual beams configuration
we compare the CPV sensitivity of the scenario with two beams and
$15\times10^{20}$ POT for C2P to that of a single beam from CERN with an
exposure of $22\times10^{20}$ POT. This is illustrated in
\reffig{fig:comparecpv}. As it is evident from
this figure, the sensitivity to CPV is better when operating two
beams. To reach a similar sensitivity in the single beam scenario,
$\sim30\times10^{20}$ POT should be collected for the C2P beam. This
confirms the phenomenological expectation that a dual beam scenario, which
combines information from two different baseline distances, will allow an
increase in precision unobtainable by an equivalent increase in exposure at a
single beam facility. 

\section{Conclusions}
\label{sec:conclusions}
The proposed LBNO experiment is the outcome of a six year program of feasibility
studies supported by the European Commission. The proposed
experiment foresees in the phase I a 20~kton underground liquid Argon
detector located at the Pyh\"asalmi mine, to be incremented to a 70~kton detector in the phase II. A
neutrino beam produced at the SPS accelerator at CERN
will allow an unambiguous determination of the
MH and give a large change to discover CP violation in the leptonic sector.

In this paper we have shown how the physics case can be further
enhanced by coupling the neutrino beam from CERN with an additional
beam from the Protvino IHEP accelerator complex. The Protvino beam has been
optimised to study CP violation and the combination of C2P and P2P will provide a $3\sigma$
measurement of $\dcp$ for the 50\% of the true values of $\dcp$ in the
phase I.
With a 70~kton detector, a better sensitivity to CPV will be reached: 
assuming the two beams with the same exposure 
a $3\sigma$ sensitivity to CP violation can be reached for
the 80\% of the true values of $\dcp$ phase-space.

In the two beam scenario, we find
that the sensitivity improvement goes beyond the simple statistical increase of
having more neutrino interactions at the Far Detector since some of
the degeneracies of the parameters involved in the neutrino
mixing can be resolved by the combination of the two beams at
different $L/E$.

We conclude that an alternative solution to explore MH and CPV in the leptonic
sector can be considered with two conventional neutrino beams with modest
powers below the 1~MW threshold. Two such beams might be easier to operate
than a single multi-MW neutrino beam, which so far has never been achieved.
A study of the dependency of the CPV
sensitivity on the systematic uncertainties and on the oscillation parameter priors
for the dual baseline experiment has been performed and we
have shown that in this configuration a discovery level for CPV can
be achieved with realistic assumptions on the beam, the detector and
the systematic errors.

\section*{Acknowledgments}

We are grateful to the European Commission for the financial support of the project through the FP7 Design Studies LAGUNA (Project Number 212343) and LAGUNA-LBNO (Project Number 284518). We would also like to acknowledge the financial supports of the Lyon Institute of Origins LabEx program (ANR-10-LABEX-66). In addition, participation of individual researchers and institutions has been further supported by funds from ERC (FP7).


\bibliography{references}
\bibliographystyle{utcaps}
\bibliographystyle{unsrt}


\bibliographystyle{apsrev4-1}
\bibliography{Prot}{}

\begin{thebibliography}{9}

\bibitem{Maki:1962mu}
  Z.~Maki, M.~Nakagawa and S.~Sakata,
  ``Remarks on the unified model of elementary particles,''
  Prog.\ Theor.\ Phys.\  {\bf 28} (1962) 870.
  
\bibitem{Pontecorvo:1967fh}
  B.~Pontecorvo,
  ``Neutrino Experiments and the Problem of Conservation of Leptonic Charge,''
  Sov.\ Phys.\ JETP {\bf 26} (1968) 984
   [Zh.\ Eksp.\ Teor.\ Fiz.\  {\bf 53} (1967) 1717].

\bibitem{An:2012eh} 
  F.~P.~An {\it et al.}  [DAYA-BAY Collaboration],
  ``Observation of electron-antineutrino disappearance at Daya Bay,''
  Phys.\ Rev.\ Lett.\  {\bf 108}, 171803 (2012)
  [arXiv:1203.1669 [hep-ex]].
    
\bibitem{Abe:2013hdq} 
  K.~Abe {\it et al.}  [T2K Collaboration],
  ``Observation of Electron Neutrino Appearance in a Muon Neutrino Beam,''
  Phys.\ Rev.\ Lett. {\bf 112} 061802 (2014) [arXiv:1311.4750 [hep-ex].

\bibitem{Stahl:2012exa} 
 A.~Stahl {\it et al.}, 
Expression of Interest for a 
``Very long baseline neutrino oscillation experiment (LBNO).''; 
CERN SPSC, June, 2012. (CERN-SPSC-2012-021 (SPSC-EOI-007)).

\bibitem{Agarwalla::2013kaa} 
  S.~K.~Agarwalla {\it et al.}  [LAGUNA-LBNO Collaboration],
  ``The mass-hierarchy and CP-violation discovery reach of the LBNO long-baseline neutrino experiment,''
  JHEP {\bf 1405}, 094 (2014)
  [arXiv:1312.6520 [hep-ph]].
  
\bibitem{Adams:2013qkq}
  C.~Adams {\it et al.}  [LBNE Collaboration],
  ``The Long-Baseline Neutrino Experiment: Exploring Fundamental Symmetries of the Universe,''
  arXiv:1307.7335 [hep-ex].
  
\bibitem{Bass:2013vcg}
  M.~Bass {\it et al.}  [LBNE Collaboration],
  ``Baseline optimization for the measurement of CP violation and mass hierarchy in a long-baseline neutrino oscillation experiment,''
  arXiv:1311.0212 [hep-ex].
  
\bibitem{Kearns:2013lea}
  E.~Kearns {\it et al.}  [Hyper-Kamiokande Working Group Collaboration],
  ``Hyper-Kamiokande Physics Opportunities,''
  arXiv:1309.0184 [hep-ex].

\bibitem{Patzak:2012rz} 
  T.~Patzak [LAGUNA-LBNO Collaboration],
  ``LAGUNA and LAGUNA-LBNO: Future megaton neutrino detectors in Europe,''
  Nucl.\ Instrum.\ Meth.\ A {\bf 695}, 184 (2012).
  
\bibitem{Rubbia:2013zqa} 
  A.~Rubbia,
  ``LAGUNA-LBNO: Design of an underground neutrino observatory coupled to long baseline neutrino beams from CERN,''
  J.\ Phys.\ Conf.\ Ser.\  {\bf 408}, 012006 (2013).
    
\bibitem{Rubbia:2009md}
  A.~Rubbia,
  ``Underground Neutrino Detectors for Particle and Astroparticle Science: The Giant Liquid Argon Charge Imaging ExpeRiment (GLACIER),''
  J.\ Phys.\ Conf.\ Ser.\  {\bf 171} (2009) 012020
  [arXiv:0908.1286 [hep-ph]].
  
\bibitem{Abe:2007bi} 
  T.~Abe {\it et al.}  [ISS Detector Working Group Collaboration],
  ``Detectors and flux instrumentation for future neutrino facilities,''
  JINST {\bf 4}, T05001 (2009)
  [arXiv:0712.4129 [physics.ins-det]].

\bibitem{Agarwalla::2014tca} 
  S.~K.~Agarwalla {\it et al.}  [LAGUNA-LBNO Collaboration],
  ``Optimised sensitivity to leptonic CP violation from spectral information: the LBNO case at 2300 km baseline,''
  arXiv:1412.0593 [hep-ph].

\bibitem{Paley:2013sta}
  J.~M.~Paley [NOvA and LBNE Collaborations],
  ``The search for CP violation and the determination of the neutrino mass hierarchy in NO?A and LBNE,''
  PoS ICHEP {\bf 2012} (2013) 393.
  
\bibitem{Coloma:2011pg}
  P.~Coloma and E.~Fernandez-Martinez,
  ``Optimization of neutrino oscillation facilities for large $\theta_{13}$,''
  JHEP {\bf 1204} (2012) 089
  [arXiv:1110.4583 [hep-ph]].

\bibitem{Arafune:1997hd} 
  J.~Arafune, M.~Koike and J.~Sato,
  ``CP violation and matter effect in long baseline neutrino oscillation experiments,''
  Phys.\ Rev.\ D {\bf 56}, 3093 (1997)
  [Erratum-ibid.\ D {\bf 60}, 119905 (1999)]
  [hep-ph/9703351].

\bibitem{Barger:2001yr}
  V.~Barger, D.~Marfatia and K.~Whisnant,
  ``Breaking eight fold degeneracies in neutrino CP violation, mixing, and mass hierarchy,''
  Phys.\ Rev.\ D {\bf 65} (2002) 073023
  [hep-ph/0112119].

\bibitem{Barabash:2002zd}
  L.~S.~Barabash, S.~A.~Baranov, Y.~A.~Batusov, S.~A.~Bunyatov, V.~Y.~Valuev, I.~A.~Golutvin, O.~Y.~Denisov and M.~Y.~Kazarinov {\it et al.},
  ``The 'IHEP-JINR Neutrino Detector' at neutrino beams of the U-70 accelerator,''
  Instrum.\ Exp.\ Tech.\  {\bf 46} (2003) 300
   [Prib.\ Tekh.\ Eksp.\  {\bf 46} (2003) 20].
   DOI: \url{http://dx.doi.org/10.1023/A:1024406219896}
  
\bibitem{omega} OMEGA Project, Nov.prob.fund.fiz. 2(9) 2010; 
\url{http://www.ihep.ru/}

\bibitem{Longhin:2012ae} 
  A.~Longhin,
  ``Optimization of neutrino beams for underground sites in Europe,''
  arXiv:1206.4294 [physics.ins-det].
  
\bibitem{Rubbia:2004tz} 
  A.~Rubbia,
  ``Experiments for CP violation: A Giant liquid argon scintillation, Cerenkov and charge imaging experiment?,''
  hep-ph/0402110.
      
\bibitem{Cervera:2010rz} 
  A.~Cervera, A.~Laing, J.~Martin-Albo and F.~J.~P.~Soler,
  ``Performance of the MIND detector at a Neutrino Factory using realistic muon reconstruction,''
  Nucl.\ Instrum.\ Meth.\ A {\bf 624}, 601 (2010)
  [arXiv:1004.0358 [hep-ex]].

\end{thebibliography}
\vspace{0.5cm}
\end{document}